\documentclass[aps,pre,amsmath,amssymb,preprint,groupedaddress]{article}
\usepackage[latin9]{inputenc}
\usepackage{geometry}
\usepackage{array}
\usepackage{float}
\usepackage{multirow}
\usepackage{graphicx}
\usepackage{setspace}

\makeatletter

\providecommand{\tabularnewline}{\\}

\newcommand{\lyxaddress}[1]{
\par {\raggedright #1
\vspace{1.4em}
\noindent\par}
}

\@ifundefined{showcaptionsetup}{}{%
 \PassOptionsToPackage{caption=false}{subfig}}
\usepackage{subfig}
\makeatother

\begin{document}
\begin{singlespace}

\title{\noindent Fast assessment of structural models of ion channels based
on their predicted current-voltage characteristics}
\end{singlespace}

\author{Witold Dyrka, Monika Kurczynska, Bogumil M. Konopka, Malgorzata Kotulska}

\maketitle

\lyxaddress{\begin{center}
Department of Biomedical Engineering, Wroclaw University of Technology,
\\
Wybrzeze Wyspianskiego 27, 50-370 Wroclaw, Poland
\par\end{center}}

\begin{center}
E-mail: \{witold.dyrka, monika.kurczynska, bogumil.konopka, malgorzata.kotulska\}@pwr.wroc.pl\\

\par\end{center}

\begin{center}
Running title: Function-oriented MQAP for ion channels
\par\end{center}
\begin{abstract}
\begin{doublespace}
Computational prediction of protein structures is a difficult task,
which involves fast and accurate evaluation of candidate model structures.
We propose to enhance single model quality assessment with a functionality
evaluation phase for proteins whose quantitative functional characteristics
are known. In particular, this idea can be applied to evaluation of
structural models of ion channels, whose main function - conducting
ions - can be quantitatively measured with the patch-clamp technique
providing the current-voltage characteristics. The study was performed
on a set of KcsA channel models obtained from complete and incomplete
contact maps. A fast continuous electrodiffusion model was used for
calculating the current-voltage characteristics of structural models.
\textit{\emph{We found that the computed charge selectivity and total
current were}} sensitive to structural and electrostatic quality of
models. In practical terms, we show that evaluating predicted conductance
values is an appropriate method to eliminate modes with an occluded
pore or with multiple erroneously created pores. Moreover, filtering
models on the basis of their predicted charge selectivity results
in a substantial enrichment of the candidate set in highly accurate
models. In addition to being a proof of the concept, our function-oriented
single model quality assessment tool can be directly applied for evaluation
of structural models of strongly-selective protein channels. Finally,
our work raises an important question whether a computational validation
of functionality should not be included in the evaluation process
of structural models, whenever possible.

\end{doublespace}

\end{abstract}
\begin{doublespace}

\section{Background}
\end{doublespace}

\begin{doublespace}
Currently there are over 48 million of protein sequences stored in
the resources of the Uniprot Consortium, while only 109 000 structures
are deposited in the Protein Data Bank (PDB) \cite{be00}, and the
gap is constantly increasing. Computational methods for protein structure
prediction are believed to be able to solve this problem. These methods
may help to identify and counteract causes of various pathological
processes through computational drug design \cite{Winter2012,Hao2012},
drug target identification \cite{Lou2010}, and protein design \cite{Koga2012}. 

Computational prediction of protein structures is a difficult task,
which also involves fast and accurate evaluation of candidate model
structures. The ultimate verification of quality of a protein model
requires availability of the native structure, or at least of its
close homologs. Typically, the assessment is based on deviations between
positions of equivalent atoms in the native protein structure and
in the assessed model. Classical methods include the Root Mean Square
Deviation (RMSD) or Global Distance Test (GDT) used in the Critical
Assessment of protein Structure Prediction competition (CASP) \cite{Moult2014}.
There are also other methods, which express structural dissimilarity
between structures, combining global and local measures or considering
only some of distances \cite{zemla99,rychlewski03,ortiz02,betancourt01,holm95,reva98,zhang04}. 

In real life situations native structures are often not attainable,
which makes model evaluation a challenge. To resolve it, numerous
Model Quality Assessment Programs (MQAPs), which estimate the quality
of produced models and select the best predictions, have been proposed.
MQAPs can be divided into three main groups: single-model, quasi single-model,
and consensus methods. Consensus methods (also known as clustering
methods) rank models in an ensemble in order to provide relative quality
scores \cite{Cheng2009,Wang2011,Skwark2013}. Quasi single-model class
include methods which evaluate a model against structural templates
\cite{Konopka2012,McGuffin2013}. Finally, single-model methods (often
referred to as \emph{true MQAPs}) predict similarity between a single
model and the unknown native structure based on a wide range of structure-
and sequence-based features of assessed models, such as solvent accessible
area, secondary structure, residue and atom contact maps, evolutionary
information, statistical potentials \cite{Wallner2006,Benkert2008,Ray12}.
The CASP10 experiment showed that consensus MQAPs outperformed single
and quasi single-model methods in case of easy and moderate targets,
however in case of difficult, free modeling targets without known
homologs, the chances were even. One of the main reasons for developing
new single and quasi single-model methods is that the consensus methods
are unable to detect low quality models if the whole ensemble of models
consists solely of low quality structures. Moreover the ability of
consensus methods to select the best models in groups of similar structures
is limited. 

In this work we propose to enhance single model assessment with a
functionality evaluation phase for proteins whose quantitative functional
characteristics are known. This approach can yield useful knowledge
showing whether a protein model is functionally correct, which is
complementary to the typical assessment based on structural features.
The main difficulty is efficient measuring and modeling the functionality.
It needs to be an experimentally measurable property that is sensitive
to structural details of a molecule. At the same time, a modeling
method needs to be fast enough to efficiently score hundreds of structural
models.

Here, we apply this idea to evaluation of structural models of ion
channels. The main function of these proteins is conducting ions,
which can be quantitatively measured with the patch-clamp technique
providing current-voltage (I-V) characteristics of a single channel
\cite{ham81}. Thus, current-voltage characteristics can be used as
a benchmark functionality for the structural model assessment. In
principle, calculation of complete I-V curve resulting from a model
structure can be performed with Molecular Dynamics (MD; \cite{ran99,imw02a}),
which treats the pore and ions in a fully discrete way, or with Brownian
Dynamics (BD; \cite{chu99,mar07}), which treats the pore and the
solute in a continuous manner and the ions discretely. However, both
methods are computationally expensive and thus slow. Especially MD
is inappropriate for prediction of the current. The alternative is
the 3-Dimensional Poisson-Nernst-Planck flow model (3D PNP), a continuous
steady-state theory, in which ions are represented by their position-dependent
average concentrations \cite{kur99,nos04,dyr08}. 3D PNP is less accurate
than MD and BD methods but manyfold faster, typically 3-5 CPU minutes
for one channel structure \cite{dyr13}. While, due to its simplicity,
the classical 3D PNP is generally not suitable to model complex physical
phenomena, it has been shown to be capable of accounting for effects
of single point mutations and of predicting I-V characteristics of
the quality sufficiently good for a MQAP \cite{fur08,dyr13}.

In this study, we apply the computationally enhanced 3D PNP model
\cite{dyr08} as a function-oriented single-model MQAP on a set of
structural models of the KcsA ion channel. First, models of diverse
quality are obtained from complete and incomplete contact maps. Then,
relations between channel structural and functional features are investigated.
Finally, the predictive power of selected functional characteristics
is assessed.
\end{doublespace}

\begin{doublespace}

\section{Materials and Methods}
\end{doublespace}

\begin{doublespace}
KcsA is a relatively well-studied potassium channel for which experimentally
solved structure in the open-conductive configuration is available
in the PDB under accession number 3FB8. It is relatively small --
its transmembrane domain consists of 4 identical units of 87 amino
acids each. Patch-clamp measurements at $\pm100$\,mV revealed relatively
high conductance from 57 to 75\,pS, mild outward rectification (1.29)
and infinite cation to anion selectivity \cite{lem01} (Tab.~\ref{tab:reference-param}). 

In order to generate a set of models of diverse quality, the experimental
structure 3FB8 was reduced to contact maps of information completeness
varying from 30\% to 100\%. Then, spatial coordinates were reconstructed
from the contact maps using \emph{C2S\_pipeline}, which applies several
state-of-the-art bioinformatic tools \cite{Konopka2014}. Structural
quality of a reconstructed model was measured using overall and single
amino acid RMSD related to the original PDB structure, the diameter
of entrance to the selectivity filter (SF) and deviation of oxygen
atoms in SF.

The electrostatics of channel models was calculated using the Poisson-Boltzmann
method and the ion flux was computed using the classical 3D PNP model.
The current-voltage characteristics were quantified at external voltage
of $\pm$100~mV using plain values and absolute deviations of the
inward and outward current (or equivalent conductances), inward and
outward charge selectivity (i.e. ratio of cation to anion current)
and rectification of the current. When applied to the reference structure,
the electrodiffusion model properly predicted outward rectification
of the channel and virtually infinite cation to anion selectivity
(above 100:1) while total currents were underestimated 3-4-fold (Tab.~\ref{tab:reference-param})
\cite{dyr13}. 
\end{doublespace}

\begin{doublespace}

\subsection{Computational pipeline}
\end{doublespace}

\begin{doublespace}

\paragraph*{Generating a contact map}
\end{doublespace}

\begin{doublespace}
Our in-house software was used to generate a contact map (CMAP) based
on the PDB file. A CMAP was a square matrix of -1, 0 and 1. A pair
of residues was assumed to be in contact if Ca atoms of both residues
were within 12~Å of one another. This distance was previously reported
as the optimal contact distance for a CMAP-based protein reconstruction
\cite{dua10}. Remaining pairs were attributed a status of non-contact
in the CMAP. In order to obtain models of different qualities CMAPs
reduced to 90\%, 70\% 50\%, and 30\% of information were also generated.
CMAP reduction was conducted by substituting the specified percentage
of randomly selected contacts and the same percentage of non-contacts
with the status of ``unknown''. The selection was conducted with
the uniform distribution, therefore equal portions of information
on contact sites were lost in all parts of the structure.
\end{doublespace}

\begin{doublespace}

\paragraph*{Modeling a channel structure from the contact map}
\end{doublespace}

\begin{doublespace}
Spatial coordinates of a channel were reconstructed from the contact
map in a three step procedure \emph{C2S\_pipeline}, which applied
several state-of-the-art bioinformatic tools \cite{Konopka2014}.
Coordinates of C$\alpha$ atoms were estimated based on constraints
imposed by the contact map using FT-COMAR \cite{vas08a,vas08b}. The
protein backbone was reconstructed by SABBAC \cite{mau06} and side-chains
were added using SCWRL \cite{kri09}. The protocol was adapted for
modeling multimeric symmetric proteins (see \cite{Konopka2014}).
The structural quality of constructed models was measured using the
following features:
\end{doublespace}
\begin{itemize}
\begin{doublespace}
\item Full model RMSD related to the original PDB structure, 
\item Model C$_{\alpha}$-C$_{\beta}$ RMSD related to the original PDB
structure,
\item RMSD of each model amino acid related to the original structure, 
\item Diameter of the selectivity filter (SF), 
\item Deviation of the selectivity filter oxygen atom, related to the original
structure. \end{doublespace}

\end{itemize}
\begin{doublespace}

\paragraph*{Modeling functional characteristics for each predicted channel structure}
\end{doublespace}

\begin{doublespace}
Two types of functional characteristics were calculated for each reconstructed
protein structure: the electrostatic profile at the pore axis, and
the current-voltage characteristics. The channel was fitted in the
129x129x129 grid at the 1~Å resolution for the Poisson-Boltzman calculations
using Adaptive Poisson-Boltzmann Solver (APBS; \cite{bak01}). Electrostatic
profiles were obtained in absence of ions and at no external voltage.
Correctness of the electrostatic profile was quantified using t\textit{\emph{he
Root Mean Square Error (RMSE) }}in reference to the profile calculated
for the original channel.

Current-voltage characteristics were determined with 3D PNP Solver
using the grids obtained from APBS. The dielectric constants were
assumed as $\epsilon$ = 4 for the protein and $\epsilon$ = 80 for
the solute. PNP calculations were carried under parametrization optimized
for narrow channels (see \cite{dyr13}), including grid spacing $\Delta$
= 2~Å, partition coefficient $\xi$ = 0.4, dielectric constant in
the pore $\epsilon$ = 40 and \textit{sphere unified} model for determining
pore-radius dependent diffusion coefficient. Computational results
obtained from 3D PNP Solver on the native channel structure were used
as the reference characteristics for assessment of predicted models.
The current-voltage characteristics were quantified at external voltage
of $\pm$100~mV using the following functional features:
\end{doublespace}
\begin{itemize}
\begin{doublespace}
\item Currents 
\end{doublespace}

\begin{itemize}
\begin{doublespace}
\item inward and outward cationic currents ($I_{in}^{+},I_{out}^{+}$), 
\item inward and outward anionic currents ($I_{in}^{-},I_{out}^{-}$), 
\item inward and outward (total) currents ($I_{in},I_{out}$); \end{doublespace}

\end{itemize}
\begin{doublespace}
\item Inward and outward charge selectivities (i.e. ratio of cation to anion
current: $I_{in}^{+}/I_{in}^{-},I_{out}^{+}/I_{out}^{-}$); 
\item Rectifications of 
\end{doublespace}

\begin{itemize}
\begin{doublespace}
\item cationic current ($|I_{out}^{+}/I_{in}^{+}|$), 
\item anionic current ($|I_{out}^{-}/I_{in}^{-}|$), 
\item (total) current ($|I_{out}/I_{in}|$). \end{doublespace}

\end{itemize}
\end{itemize}
\begin{doublespace}
Note that currents can be easily converted to conductance: 
\[
G=|I/V|,
\]
where $V$ is the electric potential applied to the membrane. The
equivalence of the current and conductance is often utilized in the
following of the document.

In addition to the plain values of the currents, selectivities and
rectification, their deviations from the current, selectivity and
rectification - calculated for the original protein structure - were
also calculated. The deviation of current was calculated as a difference:
\[
\Delta I=I_{model}-I_{reference},
\]
The deviation of charge selectivity, and the deviation of rectification
were calculated as a natural logarithm of a quotient: 
\[
\Delta(I^{+}/I^{-})=\left|ln\frac{I_{model}^{+}/I_{model}^{-}}{I_{reference}^{+}/I_{reference}^{-}}\right|,
\]
 
\[
\Delta(I_{out}/I_{in})=\left|ln\frac{|I_{out:model}/I_{in:model}|}{|I_{out:reference}/I_{in:reference}|}\right|.
\]

Note that wherever the term ``deviation'' is used throughout this
document, it always refers to the absolute deviation.

Dependencies between structural and functional features were evaluated
in terms of Kendall's $\tau$ rank correlation coefficient \cite{ken38}. 

Datasets with calculated values of functional and structural features,
and with Kendall's $\tau$ and p-values for their correlations, are
available as supplemental data (see Supplemental information 1).
\end{doublespace}

\begin{doublespace}

\subsection{Criteria of functional validity \label{sub:Criteria-of-functional}}
\end{doublespace}

\begin{doublespace}
Current-voltage characteristics obtained for the KcsA open-state conducting
structure (PDB: 3FB8) using 3D PNP Solver in our previous work \cite{dyr13}
were used to determine criteria for functional quality assessment
of predicted KcsA structures. As the 3D PNP is a semi-quantitative
model, thresholds of the functional features should not be too conservative.
In this study the following cutoffs were applied: 
\end{doublespace}
\begin{itemize}
\begin{doublespace}
\item Total inward and outward conductance at $\pm$100~mV: 
\[
G_{in},G_{out}>10~pS,
\]
which is equal to the following condition for the total inward and
outward current at $\pm$100~mV: 
\[
|I_{in}|,|I_{out}|>1~pA.
\]
Note that this threshold corresponds to roughly 1/2 of the computational
inward conductance and 2/3 of the computational outward conductance
of the original KcsA structure 3FB8 \cite{dyr13}.
\item Inward and outward cation to anion selectivity ratio were arbitrary
set to: 
\[
G_{in}^{+}/G_{in}^{-}=I_{in}^{+}/I_{in}^{-}>10:1\;\mathrm{or}\;50:1,
\]
\[
G_{out}^{+}/G_{out}^{-}=I_{out}^{+}/I_{out}^{-}>10:1\;\mathrm{or}\;50:1
\]

\item Outward rectification at $\pm$100~mV: 
\[
G_{out}/G_{in}=|I_{out}/I_{in}|>1.0.
\]
\end{doublespace}

\end{itemize}
\begin{doublespace}
The above defined thresholds provide a intuitive notion of functionally
admissible model-structures. In addition we assume that a predicted
model is \emph{conducting} when its calculated inward and outward
conductance are both within the range of 1~pS and 1~nS. The value
of 1~pS is often regarded as the bottom threshold for ionic channels
\cite{jay08}. We also found that the conductance above 1~nS is an
indicator of a porous, leaky protein structure (i.e. a structure with
multiple erroneously created pores). 
\end{doublespace}

\begin{doublespace}

\subsection{Predictive power of functional characteristics}
\end{doublespace}

\begin{doublespace}
The functional features were assessed in terms of their ability to
select models that are structurally closest to the native protein.
For this purpose, functionally correct models were regarded as properly
classified only if their general C$_{\alpha}$-C$_{\beta}$ RMSD (or
RMSE of the electrostatic profile) was below a selected threshold.
Quality of binary classification at particular threshold was evaluated
in terms of Sensitivity (\textit{Sn}), Specificity (\textit{Sp}).

\[
Sn=\frac{TP}{TP+FN}
\]

\[
Sp=\frac{TN}{TN+FP}
\]
where \textit{TP} denotes the True Positive rate, which expressed
the rate of functionally correct models that were also structurally
correct (i.e. below a selected RMSD or RMSE threshold); \textit{TN}
is a True Negative rate with functionally incorrect models that were
also structurally incorrect; \textit{FP} is the False Positive rate
with functionally correct models which were structurally incorrect;
\textit{FN} is the False Negative rate with functionally incorrect
models that were structurally correct. Additionally, Matthew's correlation
coefficient (\textit{MCC}) and Accuracy (\emph{ACC}) were also calculated:

\[
MCC=\frac{TP\cdot TN-FP\cdot FN}{\sqrt{(TP+FP)(TP+FN)(TN+FP)(TN+FN)}},
\]

\[
ACC=\frac{TP+TN}{TP+TN+FP+FN}.
\]

Overall performance of classification at various thresholds was analyzed
using the Area Under Receiver Operating Characteristic curve (\emph{AUROC})
\cite{faw06}.

The TOP100 sets consisted of 100 models which had the lowest deviations
(general C$_{\alpha}$-C$_{\beta}$ RMSD, electrostatic profile RMSE)
or the highest plain values (inward selectivity and outward selectivity)
of each feature. In case of the RMSD-based ranking, models 98th to
107th had exactly the same quality and were all included in the TOP100.
In addition to the simple rankings, two joint rankings (RMSD \& RMSE,
and inward \& outward selectivity) were generated such that both simple
rankings were extended to n models until their cross-section counted
100 models.
\end{doublespace}

\begin{doublespace}

\section{Results and Discussion}
\end{doublespace}

\begin{doublespace}

\subsection{Relation between structural and functional features}
\end{doublespace}

\begin{doublespace}

\subsubsection*{Full contact map set}
\end{doublespace}

\begin{doublespace}
In the first experiment, structures of the KcsA channel were reconstructed
based on the full contact map. The total of 430 structural models
were generated, 343 of them were \emph{conducting}, i.e. achieved
predicted conductance within the range of 1~pS and 1~nS.

All the candidate models were structurally correct as their full atom
RMSD to the original PDB structure was between 2 and 2.8~Å. However,
in terms of functionality, only 29\% of models achieved cation/anion
selectivity of 50:1 in both directions, 38\% exhibited correct outward
rectification, and 77\% achieved conductance of 10~pS in both directions.
The three functional criteria were fulfilled together by only 37 models,
which was roughly 10\% of the whole set.

\begin{figure}[H]
\centering \includegraphics[width=100mm]{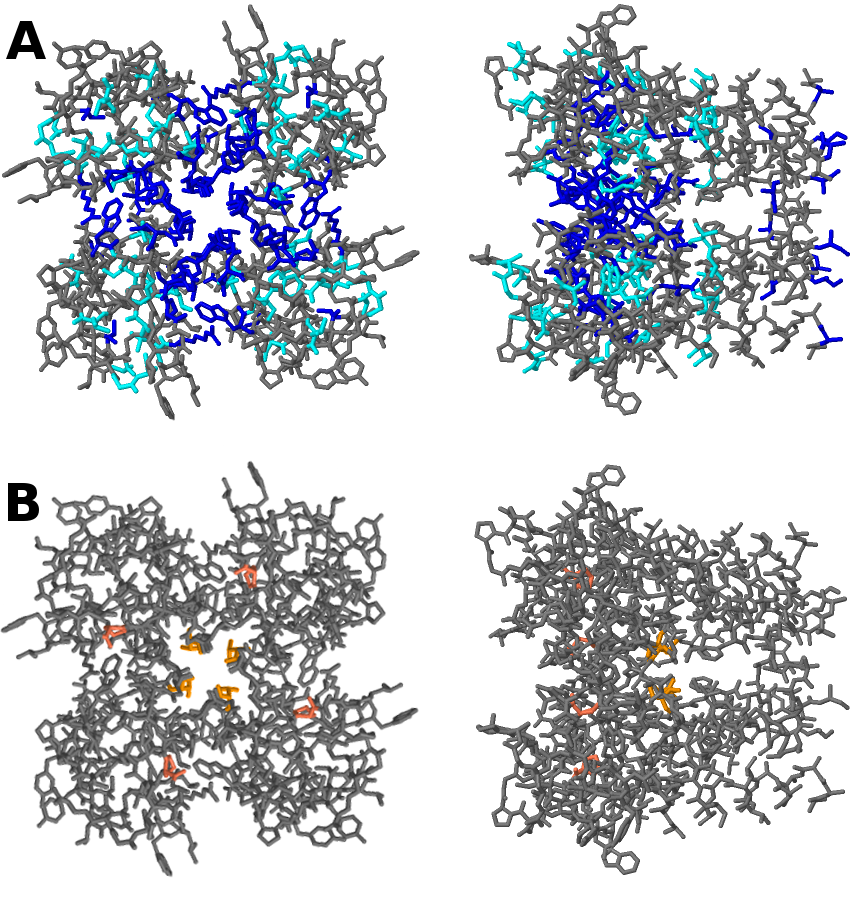} \protect\caption{Kendall's rank correlations of amino acid RMSD and deviations of functional
features in models reconstructed from full contact maps. \textbf{(A)
}All significant correlations (p-value$\leq$0.01) between an aminacid
RMSD and deviations of at least 2 (blue) or 1 (cyan) functional features.
\textbf{(B) }The strongest correlation was observed between the pore
diameter at THR75 (orange) and rectification. Other strong correlations
included the RMSD of THR75 and deviation of rectification; and the
RMSD of PRO83 (pink) and deviation of the inward anionic current.}

\label{fig:cont100_struct} 
\end{figure}

To gain more insight, Kendall's $\tau$ coefficients were calculated
between structural and functional features. The general full atom
RMSD of models correlated significantly with deviation of the inward
anionic current $\Delta I_{in}^{-}$ (Tab.~S1). Moreover, the deviation
of functional features depended on amino acids around selectivity
filter, as expected (see Fig.~\ref{fig:cont100_struct}A). The strongest
association was a positive correlation between rectification $|I_{out}/I_{in}|$
and the pore diameter at THR75, at the intracellular entrance to the
selectivity filter (p-value \textasciitilde{} 1e-10, Fig.~\ref{fig:cont100_struct}B).
Other highly significant correlations included the RMSD of THR75 and
deviation of rectification $\Delta(I_{out}/I_{in})$ and between the
RMSD of PRO83 and deviation of the inward anionic current $\Delta I_{in}^{-}$. 
\end{doublespace}

\begin{doublespace}

\subsubsection*{Reduced contact map sets\label{sub:Randomly-reduced-contact}}
\end{doublespace}

\begin{figure}[H]
\centering \includegraphics[width=150mm]{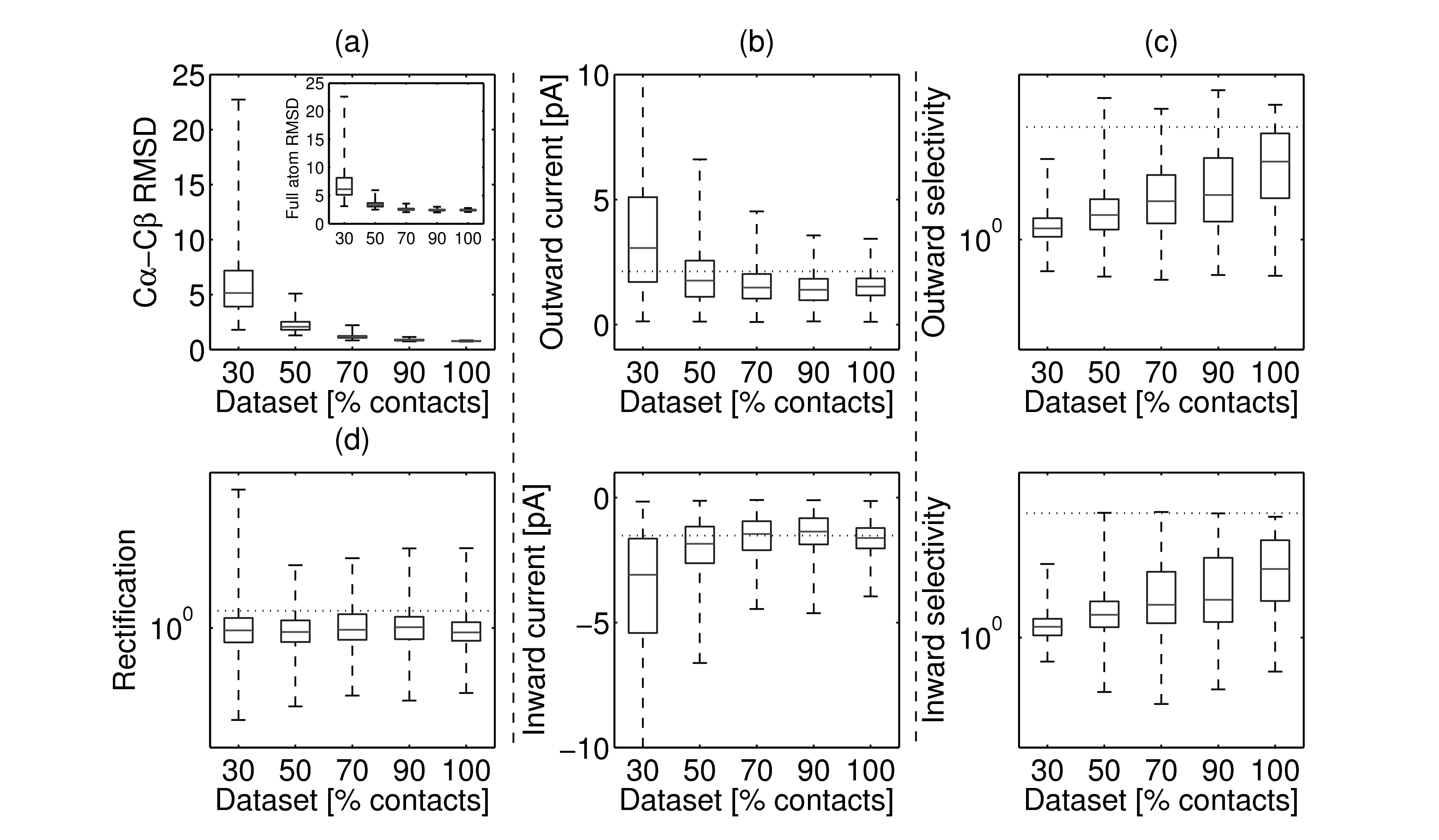} \protect\caption{Structural and functional quality of reconstructed KcsA models. \textbf{(a)
}Structural C$\alpha$-C$\beta$ (\textbf{main}) and full atom (\textbf{inset})
RMSD of predicted KcsA structures in subsets built using various percentages
of contact maps. \textbf{(b-d)} Functional characteristics of predicted
KcsA structures in 100mM KCl at $\pm$100~mV, only the\emph{ conducting
}models were considered. \textbf{(b)} total outward and inward currents,
\textbf{(c)} outward and inward cation to anion selectivity, and \textbf{(d)
}rectification (outward to inward current ratio). Notations: whiskers
- min and max, box edges - 25\% and 75\% percentile, inner line -
median, dotted line indicates value calculated for the reference structure. }

\label{fig:rmsd_by_subset} 
\end{figure}

\begin{doublespace}
In the second experiment, protein models were generated from four
randomly reduced contact map sets characterized by different information
completeness: 90\%, 70\%, 50\% and 30\%. Over 4/5 of all models achieved
full atom RMSD below 4~Å, including all models rebuilt from maps
containing 70\% or more contact information (Tab.\,S2). However,
this high RMSD threshold was reached only by $1.4\%$ of models obtained
from 30\%-complete maps. In addition, the full atom RMSD of 2/5 of
all models was below 2.5~Å. Median C$_{\alpha}$-C$_{\beta}$ RMSD
ranged from very good, i.e. 0.76~Å for full contact maps, to poor,
i.e. 6~Å for 30\%-complete maps (Fig.~\ref{fig:rmsd_by_subset}a
and Tab.\,S2). Similar pattern was observed by the full atom RMSD
(from 2.39~Å\ to 6.9~Å, respectively, Fig.~\ref{fig:rmsd_by_subset}a
(inset) and Tab.\,S2).

Functionally, the inward and outward conductance was within the range
of 1\,pS to 1~nS (\emph{conducting} models) for 1687 (78\%) models,
and exceeded 10~pS in 72\% \emph{conducting }models (Tab.~\ref{tab:reduced_contact_fun}).
The outward direction of rectification $|I_{out}|/|I_{in}|>1$ was
obtained for 41-51\% models, depending on the contact map completeness.
The median value of rectification oscillated between 0.9 and 1.0 (Fig.~\ref{fig:rmsd_by_subset}d)
and typically was significantly below the level of 1.39 calculated
for the reference structure. The inward and outward selectivity above
10:1 was reached by only 26\% predicted KcsA structures (Tab.~\ref{tab:reduced_contact_fun}),
a few models reached the inward selectivity level of the original
structure (181:1), despite relatively high randomness (Fig.~\ref{fig:rmsd_by_subset}c).
Selectivity over 50:1 was obtained for just 11\% models. Proportion
of highly selective models decreased dramatically with reduced information
in the map, for example only 2 out of 250 structures from the 30\%-complete
maps had selectivity higher than 10:1 in comparison to 193 out of
343 structures from the full map. All the functional criteria including
the selectivity above 10:1, were collectively fulfilled by 9\% of
models (almost a half of them were from the full contact maps). Only
half of them exhibited selectivity above 50:1. No structure obtained
from the 30\%-complete maps met all the functional criteria. 

\textit{\emph{General RMSDs (}}C$_{\alpha}$-C$_{\beta}$\textit{\emph{
and full atom) and deviation of charge selectivity were the most and
second most correlated pairs of model features, in terms of }}Kendall's
$\tau$ \textit{\emph{(Tab.~S3 and Fig.\,S1)}}. \textit{\emph{Deviation
of the inward current was the third most correlated functional feature
(0.21-0.24), while correlation of deviation of the outward current
was much weaker (0.12-0.15), yet still statistically significant.
}}Interestingly, correlation of deviation of the rectification with
deviation of any structural feature never exceeded range of $\tau$
between -0.09 and 0.07.
\end{doublespace}

\begin{doublespace}

\subsubsection*{Discussion.}
\end{doublespace}

\begin{doublespace}
\textit{\emph{Significant correlations between structural RMSD, and
functional deviations of the charge selectivity and the total current
(Tab.~S3), support the hypothesis that predicted structural models
could be validated on the basis of their calculated functional features
related to experimental data. }}Deviation of the anionic current (experimentally
equal to zero) was typically even more highly correlated with structural
features than the charge selectivity ($\tau$ higher up to 0.40, see
Fig.~S2), consistently with the result based on the full contact
maps (Tab.~S1). However, as experimental studies usually do not report
the anionic current, further analyzes would focus on the selectivity.
Interestingly, a significant difference in selectivity between 90\%-complete
and 100\%-complete map models suggests that this feature can be used
to distinguish between good and very good models. \textit{\emph{The
rectification}} could be, perhaps, more useful for fine tuning of
the structure, as suggested by its relatively high correlations with
some structural features in the dataset based on the full contact
maps only (Tab.~S1). However, it could be also that 3D~PNP Solver
is least suited to correctly predict rectification (see \cite{dyr13}).

The total current typically increased with decreasing completeness
of the map (Fig.~\ref{fig:rmsd_by_subset}b). This suggested a tendency
of the reconstruction pipeline to produce sparser models (with a larger
pore diameter) when information in the contact map was reduced. A
larger pore diameter could also explain why the median cation to anion
selectivity was of an order weaker for structures built from the 30\%-complete
maps than from the full maps (Fig.~\ref{fig:rmsd_by_subset}c). Indeed,
the cation to anion selectivity in the classical electrodiffusion
model applied by 3D PNP Solver is a result of presence of negative
charges in the selectivity filter and its surroundings, which prevents
passage of negatively charged ions. The effect decreases when the
selectivity filter diameter is larger than the grid resolution (here:
2~Å), as in such case, the pore is represented by two or more computational
cells in the grid and therefore negative charges in the protein wall
are partially shielded by positive charges in the solution. Thus,
the classical electrodiffusion model performs better when the pore
intersection is represented by only one computational cell -- in this
case the model is consistent with the single-file ion diffusion through
the selectivity filter \cite{hod55,koh77}. 

\begin{figure}[H]
\centering \includegraphics[width=100mm]{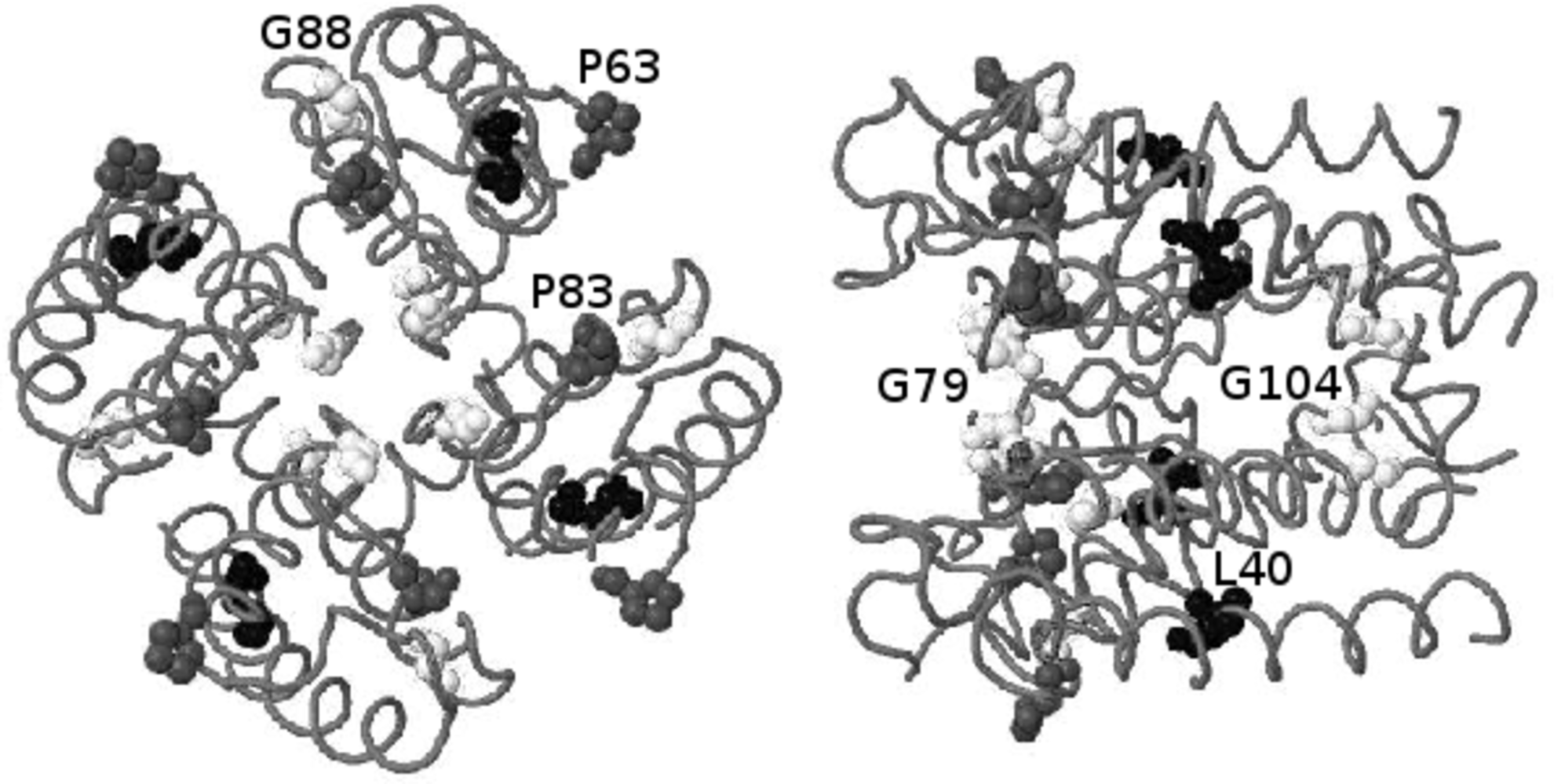} \protect\caption{Most significant Kendall's rank correlations of amino acid RMSD and
deviations of functional features in all models. Notations: black
- LEU40, gray - PRO63 (outside of the protein) and PRO83 (middle of
the protein), white - GLY79 (extracellular entrance to the SF), GLY88
(outside of the channel) and GLY104 (intracellular entrance to the
channel)}

\label{fig:cont30-100_best_corr} 
\end{figure}

The six residues, which exhibited the highest structure-function correlations
are shown in Fig.~\ref{fig:cont30-100_best_corr}. GLY79 is situated
at the extracellular entrance to the selectivity filter (SF). Therefore,
any change in its position or conformation is likely to result in
a deviation of the ion flux. ASP80 (not shown, $\tau>0.30$ only for
the oxygen position) takes part in the transformation of SF to the
non-conductive conformation \cite{cue10a}. GLY104 is located in another
region of the protein important for inactivation kinetics. It is adjacent
to PHE103, which has been recently reported to act as an interface
between the inner helical bundle and SF \cite{cue10a}. It is possible
that the RMSD measure is more sensitive to deviation of atom positions
in the small GLY104 than in the large PHE103. Prolines are typically
structurally important elements of a protein. Indeed, PRO83 is conserved
among several K$^{+}$ channels, moreover its position is next to
TYR82, another large functionally important residue (see \cite{cue10a}
supplementary information). PRO63, while quite away from SF, is adjacent
to large ARG64, one of the residues crucial for the inactivation event
\cite{Cordero-Morales2006}. A functional role of GLY88 remains unknown,
however mutations at this position were linked to disruption of tetramerization
\cite{iri02}. 
\end{doublespace}

\begin{doublespace}

\subsection{Predictive power of functional characteristics}
\end{doublespace}

\begin{doublespace}

\subsubsection*{Model classification using functional features}
\end{doublespace}

\begin{doublespace}
In this section, we examine if computed functional characteristics
of the ion flux can be used to discriminate between structurally correct
and incorrect models. In a systematic analysis, we tested various
thresholds for four functional features: deviations of inward and
outward current, and deviations of inward and outward selectivity.
The results formed the basis in a search for the optimal discriminative
values which would allow for the most reliable\textit{\emph{ model
classification}} in relation to \emph{the ground truth} given by the
general C$_{\alpha}$-C$_{\beta}$ RMSD. Only 1674 \emph{conducting}
models were considered. The C$_{\alpha}$-C$_{\beta}$ RMSD\emph{
}thresholds were fixed at values of 1~Å\ (highly accurate) and 3~Å
(correct). The optimal thresholds were selected according to the maximum
product of sensitivity (\textit{Sn}\textit{\emph{)}} and \textit{\emph{specificity}}\textit{
}\textit{\emph{(}}\textit{Sp}). Classification quality was evaluated
also in terms of accuracy (\emph{ACC}) and Matthew's correlation coefficient
(\textit{MCC}) at the optimal threshold, and in terms of the area
under the ROC curve (\emph{AUROC}) as a summary measure over all thresholds
(Tab.\ref{tab:RMSD_dev})\textit{.} 

\begin{figure}[H]
\centering\includegraphics[width=120mm]{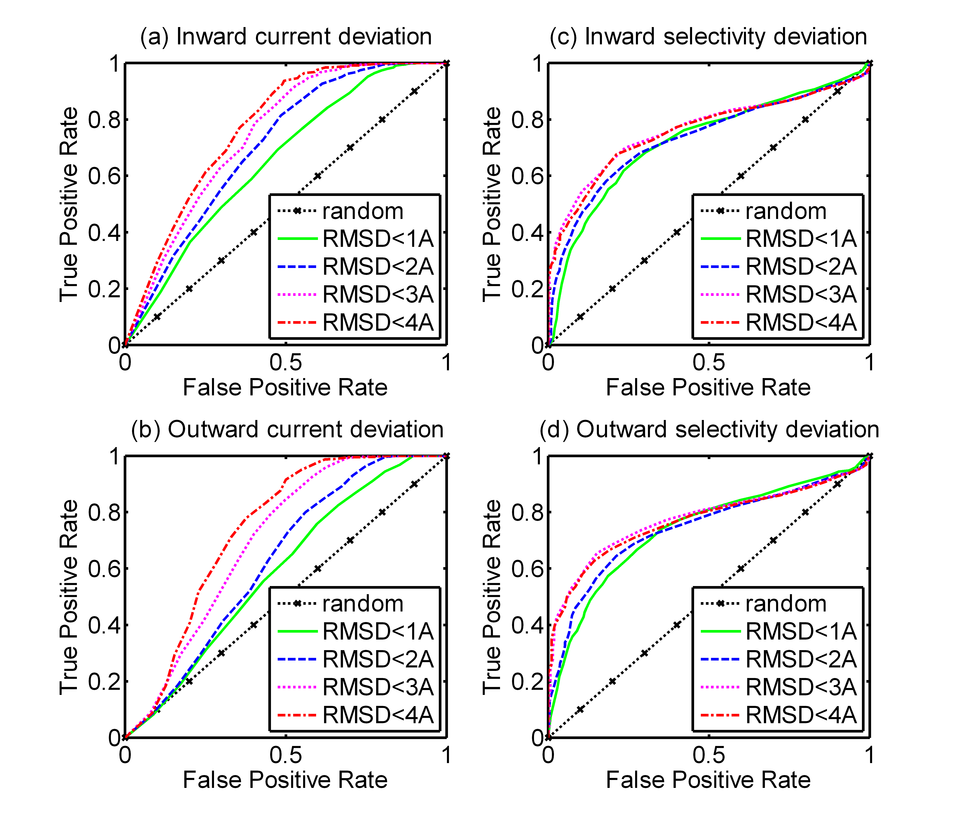} 

\protect\caption{ROC curves of model classification based on deviations of current
and selectivity at four thresholds of the general C$_{\alpha}$-C$_{\beta}$
RMSD. Only the\emph{ conducting }models were considered. The RMSD
thresholds corresponded to the following positive/negative ratios:
1\,Å:\,752/935, 2\,Å:\,1229/458, 3\,Å:\,1412/275, 4\,Å:\,1499/188.}

\label{fig:ROC_RMSD} 
\end{figure}

The classification based on deviation of the inward selectivity produced
ROC curves which were well above the diagonal for both RMSD thresholds
(\emph{AUROC} $0.72-0.78$) and shifted towards specificity (Fig.~\ref{fig:ROC_RMSD}).
The classification based on the current deviation resulted in similar
\emph{AUROC} for the RMSD \textit{\emph{of 3}}\textit{~}Å, while
\textit{\emph{it was lower for the 2~}}Å\textit{\emph{ threshold
(}}\emph{AUROC}\textit{\emph{ $0.59-0.65$)}}. In this cases the ROC
curve was shifted towards sensitivity. The deviation of selectivity
displayed the best balance between retaining good quality models and
rejecting structurally incorrect models (\textit{MCC $0.38-0.39$}).
Overall performance of the classification was better for higher RMSD
thresholds. The optimal thresholds were relatively lower for the outward
direction of selectivity (by 14-18\%), and for the inward direction
of current (by 8-20\%). In practical terms, applying the optimal thresholds
of selectivity retained ca.\,70\% of structurally correct models
(518-548 out of 752 models with RMSD\,$<1$\,Å, and 940-1015 out
of 1412 models with RMSD\,$<3$\,Å) at the cost of retaining 15-33\%
of structurally incorrect models in the group of functionally correct
models (288-313 out of 935 models with RMSD\,$\geq1$\,Å, and 42-69
out of 275 models with RMSD\,$\geq3$\,Å). Using the optimal thresholds
of current deviations as a classifier resulted in retaining 71-82\%
of correct models and 41-66\% of incorrect models.
\end{doublespace}

\begin{doublespace}

\subsubsection*{\textit{Electrostatic RMSE as a complementary ground truth}}
\end{doublespace}

\begin{figure}[H]
\subfloat[]{

\includegraphics[height=60mm]{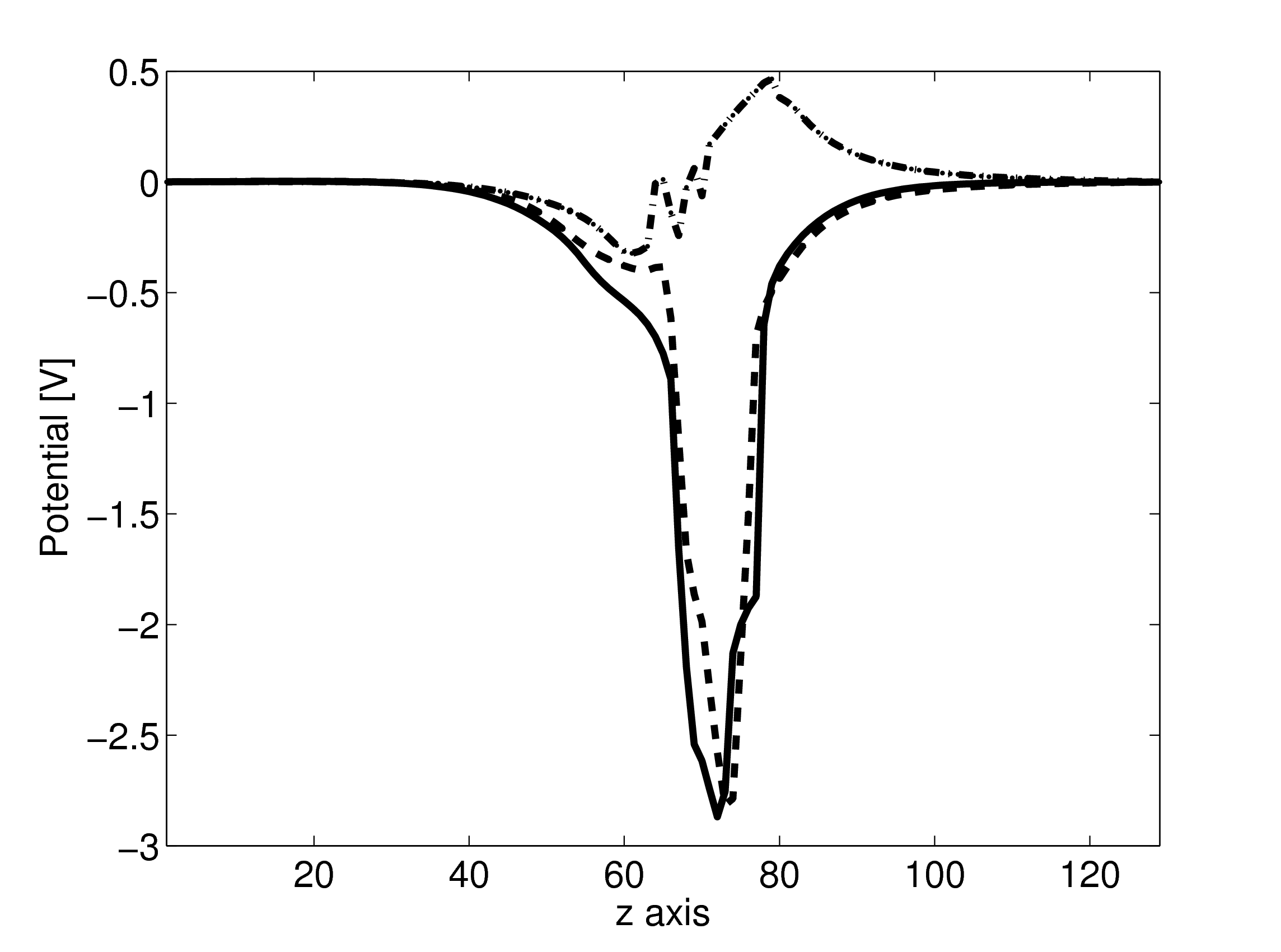}}\hfill{}\subfloat[]{

\includegraphics[height=60mm]{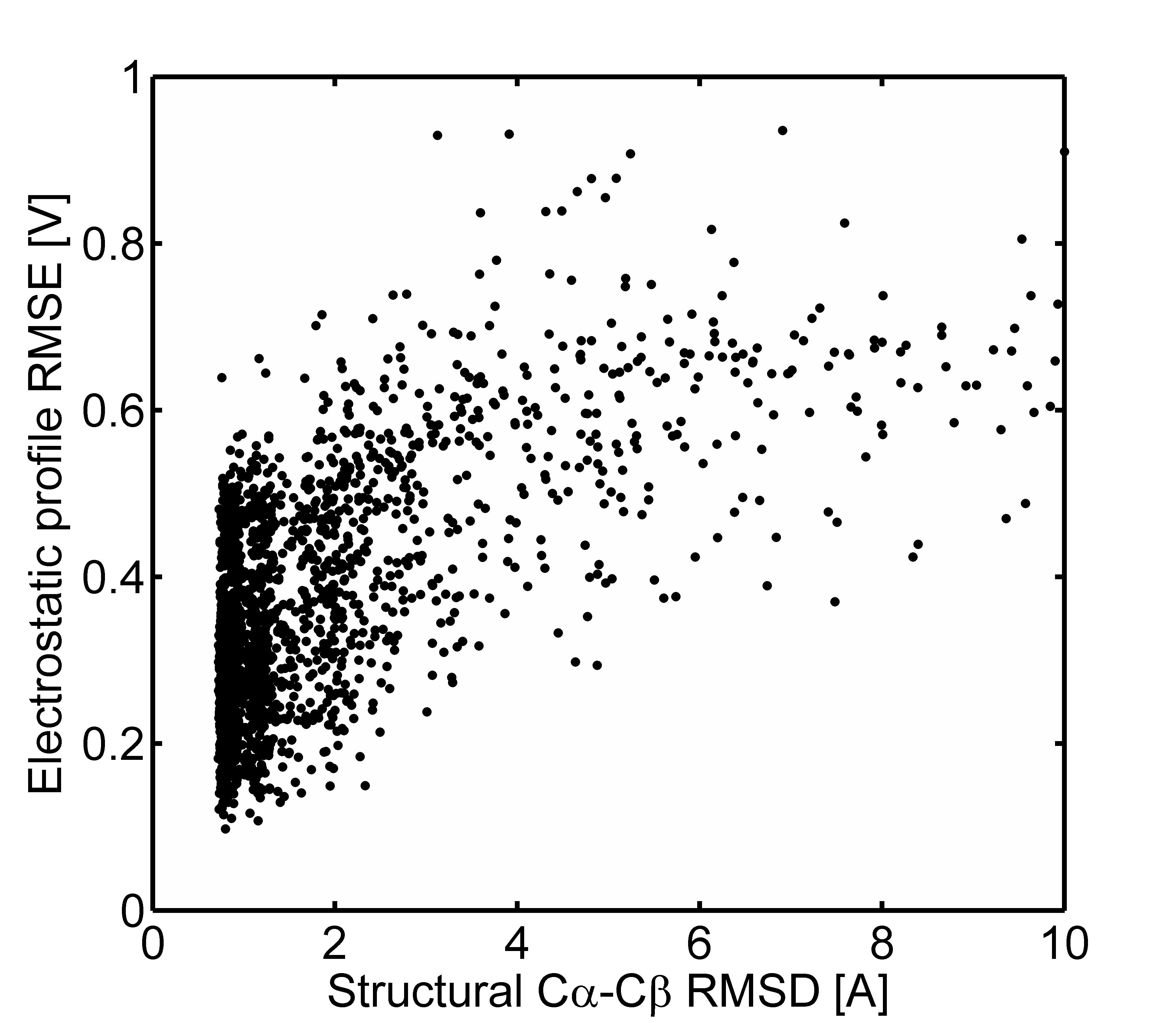}}

\protect\caption{Electrostatic profile RMSE as a complementary \emph{ground truth}.
\textbf{(a)} Exemplary electrostatic profiles of the reference channel
pore (solid) and two modeled channel pores: correct (dashed line)
and incorrect (dash-dotted line). \textbf{(b) }Scatter plot of the
electrostatic profile RMSE versus the general C$_{\alpha}$-C$_{\beta}$
RMSD. The axes have been cut at the 10\,Å and 1\,V thresholds of
RMSD and RMSE, respectively. Both measures are overally well correlated
(Kendall's $\tau=0.45$), the relation is much weaker for low RMSD
structures ($\tau=0.12$ for RMSD\,$<$\,1.7\,Å).}

\label{fig:RMSE_as_GT} 
\end{figure}

\begin{doublespace}
In section \ref{sub:Randomly-reduced-contact}, we reported that a
significant proportion of models were non-\emph{conducting} while
having a relatively low general RMSD. This raised doubts whether general
RMSD was an appropriate solitary measure of the channel structure
quality. Therefore, we propose the electrostatic potential profile
in the native channel structure as a new \emph{ground truth} (Fig.~\ref{fig:RMSE_as_GT}a),
and its Root Mean Square Error (RMSE) as an alternative to the entirely
structure-based RMSD. It can be argued that the RMSE of the electrostatic
potential profile is a measure that balances the structural and functional
quality of the channel model as the electrostatic potential profile
is determined by the structure, and determines the channel function
at the same time. While the electrostatic potential profile cannot
be measured experimentally, it could be used to assess the relationship
between structural and functional quality, and to evaluate the predictive
power of calculated current-voltage characteristics. 

First, we established the relation between the structural RMSD and
the electrostatic RMSE threshold (Fig.~\ref{fig:RMSE_as_GT}b). We
found that the electrostatic profile RMSE and the general C$_{\alpha}$-C$_{\beta}$
RMSD were generally well correlated. (Kendall's $\tau=0.45$). However,
the relation was much weaker for low RMSD structures ($\tau=0.12$
for RMSD $<$ 1.7\,Å). In this group consisting of the most accurate
models the two characteristics were complementary to each other.

\begin{figure}[H]
\centering\includegraphics[width=120mm]{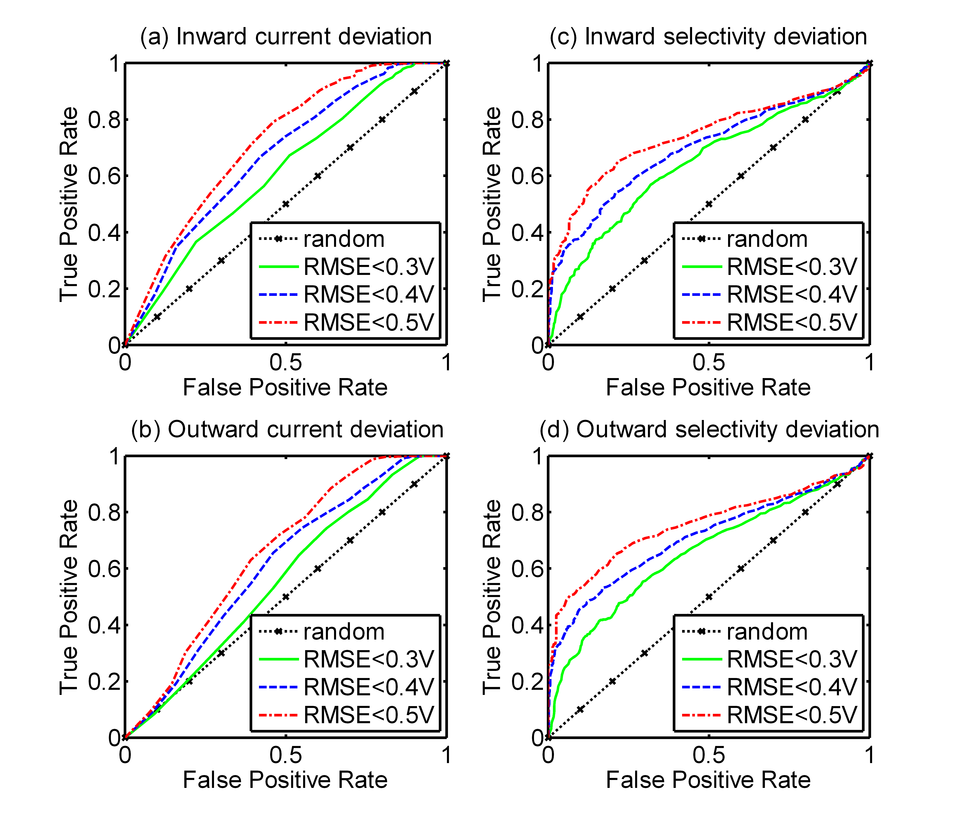} 

\protect\caption{ROC curves of model classification based on deviations of current
and selectivity at three threshold of the electrostatic profile RMSE.
Only the\emph{ conducting }models were considered. The RMSE thresholds
corresponded to the following positive/negative ratios: 0.3\,V:\,625/1049,
0.4\,V:\,1025/649, 0.5\,V:\,1324/350. The AUROC was in the following
ranges: (a) 0.61-0.72, (b) 0.56-0.66, (c) 0.64-0.74, (d) 0.66-0.76.}

\label{fig:ROC_RMSE} 
\end{figure}

Next, we searched for the optimal thresholds for the four functional
features (the inward and outward, current and selectivity deviations)
to obtain the most reliable\emph{ }\textit{\emph{classification in
terms of specificity and sensitivity product,}} related to the electrostatic
RMSE at fixed thresholds of 0.3, 0.4 and 0.5\,V. Again, only the
\emph{conducting} models were considered. Generally, classification
characteristics were similar as in the case of classification in relation
to the structural RMSD (Fig.~\ref{fig:ROC_RMSE} and Tab.\,S4).
Interestingly, overall performance of the classification was more
sensitive to changing the threshold of \emph{the ground truth} than
in the RMSD-related experiment. This finding is consistent with presumably
closer relationship between the current-voltage characteristics and
the electrostatic profile.
\end{doublespace}

\begin{doublespace}

\subsubsection*{Practical scenarios\label{sub:Practical-scenarios}}
\end{doublespace}

\begin{doublespace}
In this section, two practical scenarios of our model quality assessment
approach are analyzed. In order to emulate the-real-life use cases
plain values of functional features were used instead of deviations
related to the true structure.
\end{doublespace}

\begin{doublespace}

\paragraph*{Enriching a candidate set in high accuracy models}
\end{doublespace}

\begin{doublespace}
In the first scenario, the goal was to reduce a collection of candidate
structural models. They were subjected to criteria of functional correctness
in reference to available experimental current-voltage characteristics.
The criteria were rather liberal, keeping in mind the semi-quantitative
character of the 3D PNP model. We checked how applying intuitive functional
conditions (defined in Sec\,\ref{sub:Criteria-of-functional}) reduces
the dataset and enriches it in structures with low general C$_{\alpha}$-C$_{\beta}$RMSD
and profile RMSE (Tab.\,\ref{tab:Quality-enrichment}\,top).

\begin{figure}[H]
\protect\caption{\label{fig:Quality-enrichment}Quality enrichment of the candidate
subsets using several functional criteria. \textbf{(a)} Boxplots of
the structural C$_{\alpha}$-C$_{\beta}$ RSMD and the electrostatic
profile RMSE in groups of models fulfilling functional conditions:
\emph{cond}: \emph{the conducting models} or 1\,pS$<$$G$$<$1~nS,
C10: $G$$>$10~pS, S10: $G^{+}/G^{-}>10:1$, S50: $G^{+}/G^{-}>50:1$,
RO: $G_{out}/G_{in}>1$. Notations: whiskers - min and max, box edges
- 25\% and 75\% percentile, inner line - median. \textbf{(b) }Scatter
plot of the electrostatic profile RMSE versus the general C$_{\alpha}$-C$_{\beta}$
RMSD. The axes have been cut at the 10\,Å and 1\,V thresholds of
RMSD and RMSE, respectively. Notations: magenta - the \emph{conducting
}models\emph{ }with inward and outward selectivities $G^{+}/G^{-}>10:1$,
black - all other models.}

\subfloat[]{

\includegraphics[height=65mm]{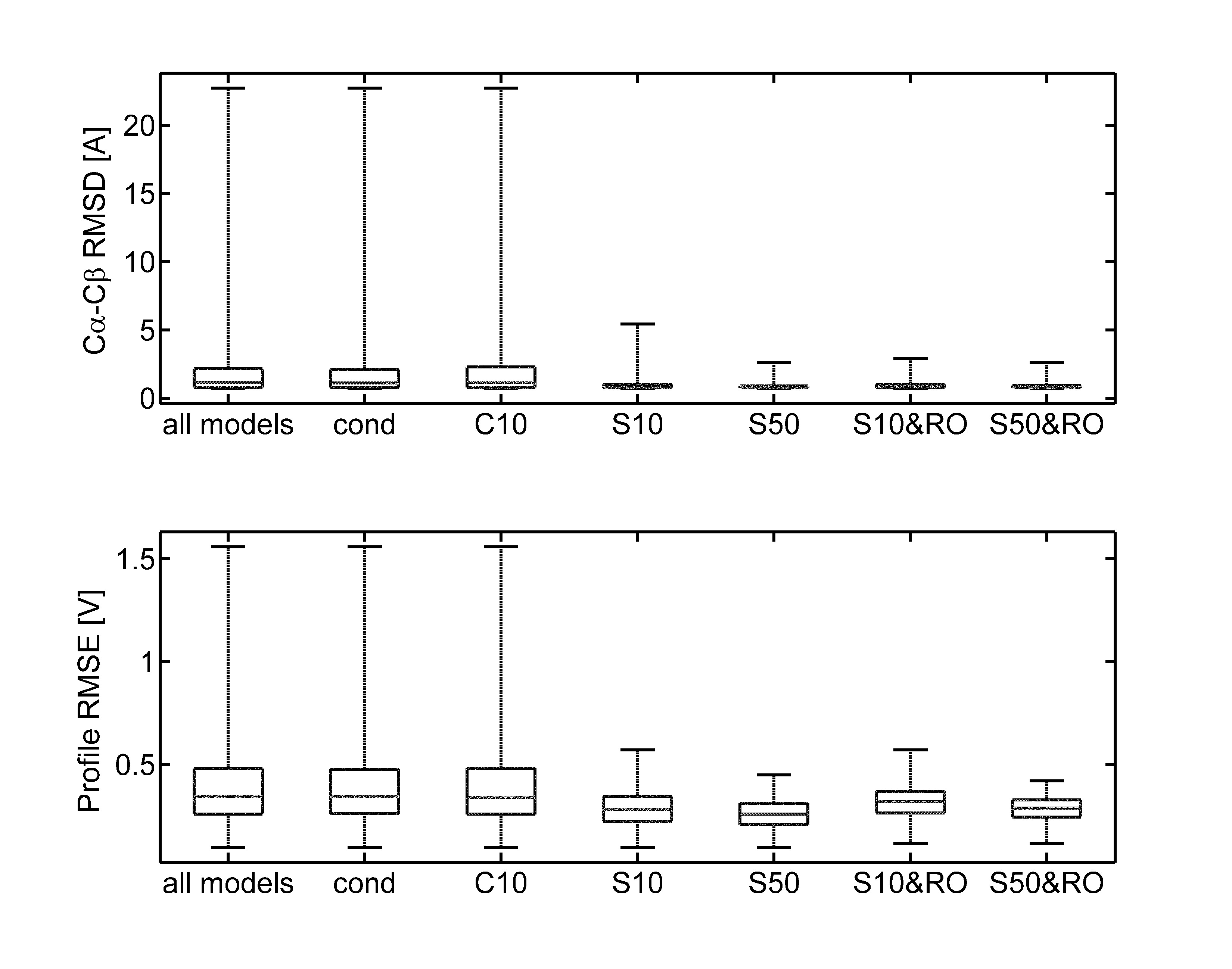}}\hfill{}\subfloat[]{

\includegraphics[height=65mm]{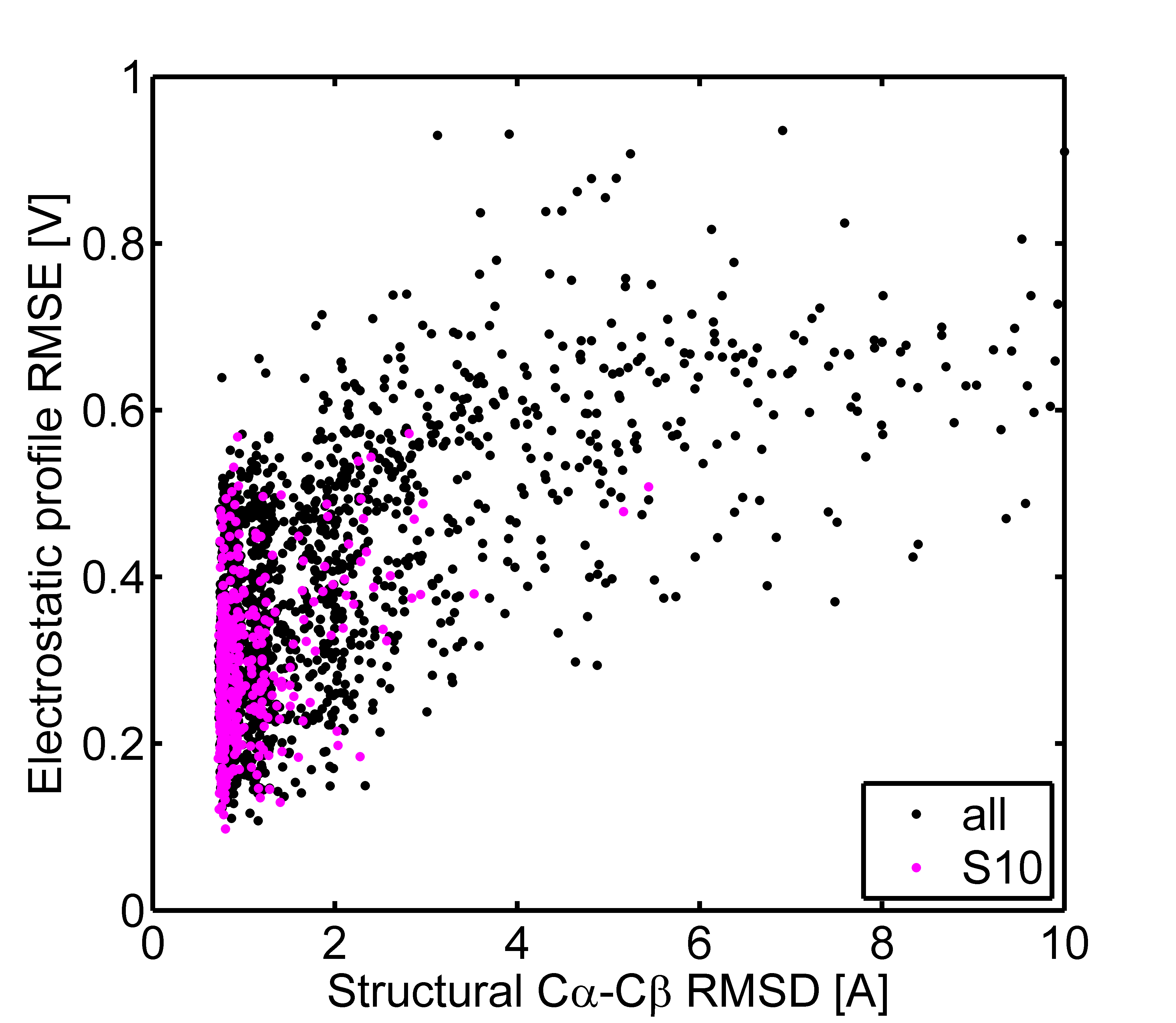}}
\end{figure}

None of conductance conditions improved quality of the resulting subset
in comparison to the initial dataset (Fig.\,\ref{fig:Quality-enrichment}a).
Unlike that, enrichment in high quality candidates due to the selectivity
criteria was substantial (Fig.\,\ref{fig:Quality-enrichment}b).
Virtually all structures with selectivity ratio above 10:1 were within
RMSD$<3$\,Å and RMSE$<$0.5\,V from the real structure in comparison
to ca. 80\% in the whole population. Moreover, fraction of highly
accurate structures increased from 44\% to 75\% (RMSD$<1$\,Å) or
from 38\% to 57\% (RMSE$<$0.3\,V). The improvement was even more
pronounced with the more stringent selectivity criteria (increase
of highly accurate fraction by almost 90\%). Median RMSD and RMSE
were reduced by 20-30\% and all exceptionally wrong models were filtered
out (no RMSD/RMSE was above $5.44$\,Å/0.58\,V for S10 or above
2.61\,Å/0.45\,V for S50, Fig.\,\ref{fig:Quality-enrichment}a).
Finally, adding the outward rectification condition slightly worsened
the candidate set in terms of enrichment in structurally and electrostatically
accurate models (Fig.\,\ref{fig:Quality-enrichment}a).
\end{doublespace}

\begin{doublespace}

\paragraph*{Selecting the best models}
\end{doublespace}

\begin{doublespace}
In the second scenario, the goal was to select the best 100 candidate
models (TOP100). We checked to what extent the 100 best models in
terms of the cation to anion selectivity overlap with the 100 best
models in terms of the general C$_{\alpha}$-C$_{\beta}$ RMSD and/or
the profile RMSE (see Tab.\,\ref{tab:Quality-enrichment}\,bottom).
Models were ranked separately in four simple categories: general C$_{\alpha}$-C$_{\beta}$
RMSD, profile RMSE, inward selectivity, outward selectivity, and in
two joint categories: RMSD \& RMSE and inward \& outward selectivity
(See Methods). In TOP100{[}RMSD{]} the RMSD ranged from 0.717 to 0.749\,Å,
in TOP100{[}RMSE{]} the RMSE ranged from 0.098 to 0.171\,V, and in
TOP100{[}RMSD\&RMSE{]} the RMSD ranged from 0.717 to 0.774\,Å and
the RMSE ranged from 0.121 to 0.232\,V. In TOP100\emph{ }{[}$\frac{I_{in}^{+}}{I_{in}^{-}}${]}
the inward selectivity ranged from 70:1 to 192:1, , in TOP100{[}$\frac{I_{out}^{+}}{I_{out}^{-}}${]}
the outward selectivity ranged from 108:1 to 518:1, and in TOP100{[}$\frac{I_{in}^{+}}{I_{in}^{-}}$\&$\frac{I_{out}^{+}}{I_{out}^{-}}${]}
the inward selectivity ranged from 62:1 to 185:1 and the outward selectivity
ranged from 89:1 to 518:1.

Probability of finding a TOP100 model from the \emph{ground-truth}-based
ranking by chance was less than 5\%. The odds increased drastically
when only the 100 most cation selective models were considered. Most
notably, the TOP100 according to outward selectivity included 33 out
of the best 100 models in terms of the joint RMSD and RMSE criterion
(7 times better than random). Enrichment in the TOP100 based on the
inward selectivity was weaker but still significant (from 1.9 times
for TOP100{[}RMSD{]} to 4.1 times for TOP100{[}RMSD\&RMSE{]}). Enrichment
in the TOP100 based on the joint ranking of inward and outward selectivity
ranged from 2.6 to 4.8 times, depending on the \emph{ground truth}.
\end{doublespace}

\begin{doublespace}

\subsubsection*{Discussion}
\end{doublespace}

\begin{doublespace}
Fairly good \emph{AUROC} values for classification based on the selectivity
\textit{\emph{and current deviations showed that the features}}\emph{
}are sensitive to structural and electrostatic quality of models and
therefore are suitable for separating models with low and high structural
RMSD or electrostatic profile RMSE. However, ranges of current defined
by the optimal deviation thresholds were below experimental values
of the current (Tab.\,\ref{tab:reference-param} and Tab.\,\ref{tab:RMSD_dev}).
In addition, we found that liberal thresholds of current, taking into
account approximate accuracy of the classical 3D PNP model, could
not be effectively used to filter out structurally or electrostatically
inaccurate models. Consequently, with the classical electrodiffusion
model, the current-based criterion can be employed only for eliminating
models with an occluded pore or with multiple erroneously created
pores. 

The optimal thresholds for selectivity deviation translate to selectivity
cutoff ranging from 2.3:1 (inward selectivity, RMSD$<$3\,Å) to from
4.1:1 (outward selectivity, RMSD$<$1\,Å). These cutoff were severely
underestimated in reference to experimental data (no anionic current)
and to computational results for the original structure (Tab.\,\ref{tab:reference-param}).
However, due to very good specificity of the selectivity-based classification
(i.e. retrieving ca. 20\% of structurally accurate models with a few
false positives, Fig.\,\ref{fig:ROC_RMSD}cd), the condition of high
selectivity (above 10:1) proved to be practical for model quality
assessment (Tab.\,\ref{tab:Quality-enrichment} and Fig.\,\ref{fig:Quality-enrichment}).
While it requires further studies to verify if the semi-quantitative
accuracy of the classical 3D PNP in predicting selectivity is sufficient
for assessment of candidate models of mildly-selective channels (such
as alpha-hemolysin, GLIC, etc.), the present study showed that the
method is appropriate for the class of strongly-selective channels.
\end{doublespace}

\begin{doublespace}

\section{Conclusions}
\end{doublespace}

\begin{doublespace}
In this study, we proposed a novel function-oriented approach to the
single model quality assessment which is complementary to existing
methods. The approach is applicable to analysis of structural models
of proteins whose quantitative functional characteristics are known.
This general idea was applied to quality assessment of structural
models of potassium channel KcsA generated from contact maps of varying
quality. The evaluation was based on current-voltage characteristics
computed for predicted structures using the classical 3D Poisson-Nernst-Planck
model, which were compared to available results from patch-clamp experiments.

We found that structural quality of candidate models, in terms of
RMSD to the original structure, was significantly correlated with
predicted conductance and charge selectivity (Kendall's rank correlation
up to 0.4). \textit{\emph{This supported the initial hypothesis that
predicted structural models could be validated on the basis of their
calculated functional features related to experimental data. }}It
was further confirmed by good performance in separating models with
low and high RMSD on the basis on deviation of current and selectivity
from their values computed for the true structure (\emph{AUROC} up
to 0.78). 

In practical terms, our approach had to deal with limitations of the
classical 3D PNP, which is a fast but approximate method and could
not accurately reproduce experimental characteristics for the reference
structure. Therefore, cutoff thresholds for assessing functional correctness
of a model had to be set liberally. Under these conditions, we showed
that evaluating predicted conductance was an appropriate method to
eliminate modes with an occluded pore or with multiple erroneously
created pores. In addition, filtering models on the basis of their
predicted charge selectivity resulted in a substantial enrichment
of the candidate set in highly accurate models. E.g. by demanding
the charging selectivity above 10:1, we obtained a high accuracy subset
containing 21\% candidate models of which 99\% had C$_{\alpha}$-C$_{\beta}$
RMSD below 3\,Å. This shows that the method can be directly applied
for evaluation of structural models of at least strongly-selective
protein channels. Moreover, it can be expected that efficiency of
our model quality assessment method will improve when more accurate
and comparably fast continuous models of ion flow in a protein channel
are available.

Our work raises an important question how to define correctness of
an ion channel model. Is the general RMSD an appropriate \emph{ground
truth} measure in this context? We found that a significant proportion
of models were occluded while having a low general RMSD. It is unlikely
that this could be uniquely attributed to the coarse resolution and
discretization used in the PNP calculations. In addition, an important
variation of electrostatic profiles was found in a group of models
characterized by C$_{\alpha}$-C$_{\beta}$ RMSD below 1\,Å. Therefore,
we investigated using the electrostatic potential profile of the reference
structure as a complementary \emph{ground truth}. Not surprisingly,
models with low and high RMSE of the electrostatic profile were well
separated on the basis of deviation of current and selectivity (\emph{AUROC}
up to 0.76). Very interestingly, the selection of 100 best models
in terms of the selectivity was significantly more enriched in TOP100
models with the the joint lowest RMSE and RMSD than in models with
the lowest RMSD. While the electrostatic profile cannot be measured
experimentally, our results indicate that predicted current-voltage
characteristics convey information about electrostatics. This important
information about correctness of a model is complementary to the general
RMSD. This suggests that, perhaps, the computational validation of
functionality should be included in the evaluation process of structural
models whenever possible.
\end{doublespace}

\begin{doublespace}

\section*{Acknowledgments}
\end{doublespace}

\begin{doublespace}
This research was partially supported by National Science Center grant
no N N519 643540 for M.Kotulska, scholarship START of Foundation for
Polish Science for W.Dyrka and from the budget for science in 2012-2015
as a research project within the program ``Diamond Grant'' DI2011
002141 for M.Kurczynska. Some of the calculations have been performed
in Wroclaw Center for Networking and Supercomputing.

\bibliographystyle{elsarticle-num}
\bibliography{PNP-MQAP_ArXiv}

\clearpage{}
\end{doublespace}

\begin{doublespace}

\section*{Figures captions}
\end{doublespace}
\begin{description}
\begin{doublespace}
\item [{Figure}] \textbf{1: Kendall's rank correlations of amino acid RMSD
and deviations of functional features in models reconstructed from
full contact maps. (A) }All significant correlations (p-value$\leq$0.01)
between an amino acid RMSD and deviations of at least 2 (blue) or
1 (cyan) functional features. \textbf{(B) }The strongest correlation
was observed between the pore diameter at THR75 (orange) and rectification.
Other strong correlations included the RMSD of THR75 and deviation
of rectification; and the RMSD of PRO83 (pink) and deviation of the
inward anionic current.
\item [{Figure}] \textbf{2: Structural and functional quality of reconstructed
KcsA models.} \textbf{(a) }Structural C$\alpha$-C$\beta$ (\textbf{main})
and full atom (\textbf{inset}) RMSD of predicted KcsA structures in
subsets built using various percentages of contact maps. \textbf{(b-d)}
Functional characteristics of predicted KcsA structures in 100mM KCl
at $\pm$100~mV, only the\emph{ conducting }models were considered.
\textbf{(b)} total outward and inward currents, \textbf{(c)} outward
and inward cation to anion selectivity, and \textbf{(d) }rectification
(outward to inward current ratio). Notations: whiskers - min and max,
box edges - 25\% and 75\% percentile, inner line - median, dotted
line indicates value calculated for the reference structure. 
\item [{Figure}] \textbf{3: Most significant Kendall's rank correlations
of amino acid RMSD and deviations of functional features in all models.}
Notations: black - LEU40, gray - PRO63 (outside of the protein) and
PRO83 (middle of the protein), white - GLY79 (extracellular entrance
to the SF), GLY88 (outside of the channel) and GLY104 (intracellular
entrance to the channel)
\item [{Figure}] \textbf{4: ROC curves of model classification based on
deviations of current and selectivity at four thresholds of the general
}C$_{\alpha}$-C$_{\beta}$\textbf{ RMSD.} Only the\emph{ conducting
}models were considered. The RMSD thresholds corresponded to the following
positive/negative ratios: 1\,Å:\,752/935, 2\,Å:\,1229/458, 3\,Å:\,1412/275,
4\,Å:\,1499/188.
\item [{Figure}] \textbf{5: Electrostatic profile RMSE as a complementary
}\textbf{\emph{ground truth}}\textbf{.} \textbf{(a)} Exemplary electrostatic
profiles of the reference channel pore (solid) and two modeled channel
pores: correct (dashed line) and incorrect (dash-dotted line). \textbf{(b)
}Scatter plot of the electrostatic profile RMSE versus the general
C$_{\alpha}$-C$_{\beta}$ RMSD. The axes have been cut at the 10\,Å
and 1\,V thresholds of RMSD and RMSE, respectively. Both measures
are overall well correlated (Kendall's $\tau=0.45$), the relation
is much weaker for low RMSD structures ($\tau=0.12$ for RMSD\,$<$\,1.7\,Å).
\item [{Figure}] \textbf{6: ROC curves of model classification based on
deviations of current and selectivity at three threshold of the electrostatic
profile RMSE.} Only the\emph{ conducting }models were considered.
The RMSE thresholds corresponded to the following positive/negative
ratios: 0.3\,V:\,625/1049, 0.4\,V:\,1025/649, 0.5\,V:\,1324/350.
The AUROC was in the following ranges: (a) 0.61-0.72, (b) 0.56-0.66,
(c) 0.64-0.74, (d) 0.66-0.76.
\item [{Figure}] \textbf{7: Quality enrichment of the candidate subsets
using several functional criteria}. \textbf{(a)} Box plots of the
structural C$_{\alpha}$-C$_{\beta}$ RSMD and the electrostatic profile
RMSE in groups of models fulfilling functional conditions: \emph{cond}:
\emph{the conducting models} or 1\,pS$<$$G$$<$1~nS, C10: $G$$>$10~pS,
S10: $G^{+}/G^{-}>10:1$, S50: $G^{+}/G^{-}>50:1$, RO: $G_{out}/G_{in}>1$.
Notations: red line - median, box edges - $25$\textsuperscript{th}
and $75$\textsuperscript{th} percentile, whiskers - min and max.
\textbf{(b) }Scatter plot of the electrostatic profile RMSE versus
the general C$_{\alpha}$-C$_{\beta}$ RMSD. The axes have been cut
at the 10\,Å and 1\,V thresholds of RMSD and RMSE, respectively.
Color code: green - the \emph{conducting }models\emph{ }with inward
and outward selectivities $G^{+}/G^{-}>10:1$, blue - remaining \emph{conducting
}models\emph{ }(1\,pS$<$$G$$<$1~nS), black - all other models.\end{doublespace}

\end{description}
\begin{doublespace}

\section*{\newpage{}Tables}
\end{doublespace}

\begin{doublespace}
\begin{table}[H]
\begin{tabular}{lrr}
\hline 
Parameter  & Experimental  & Computational \tabularnewline
\hline 
\textit{Inward (-100~mV)}  &  & \tabularnewline
\quad{}\quad{}Total conductance $I_{in}$  & 57~pS  & 15~pS \tabularnewline
\quad{}\quad{}Cation/anion selectivity $I_{in}^{+}/I_{in}^{-}$  & $\infty$:1  & 181:1 \tabularnewline
\textit{Outward (+100~mV)}  &  & \tabularnewline
\quad{}\quad{}Total conductance $I_{out}$  & 75~pS  & 21~pS \tabularnewline
\quad{}\quad{}Cation/anion selectivity $I_{in}^{+}/I_{in}^{-}$  & $\infty$:1  & 111:1 \tabularnewline
Rectification $|I_{out}/I_{in}|$  & 1.29  & 1.39 \tabularnewline
\hline 
\end{tabular}

\protect\caption{\label{tab:reference-param}\textbf{Selected experimental \cite{lem01}
and computational \cite{dyr13} parameters of KcsA I-V curves.} Computational
results obtained using 3D PNP Solver on the 3FB8 structure.}
\end{table}

\end{doublespace}

\begin{doublespace}

\subsection*{\newpage{}}
\end{doublespace}

\begin{doublespace}
\begin{table}[H]
\begin{tabular}{>{\raggedleft}p{1.5cm}>{\raggedleft}p{2cm}rrrrrr}
\hline 
\multirow{2}{1.5cm}{CMAP } & \multirow{2}{2cm}{Conducting models} & Conductance  & Rectification  & \multicolumn{2}{c}{Selectivity} & \multicolumn{2}{c}{All criteria incl.}\tabularnewline
 &  & $G>$10\,pS & $|I_{out}|>|I_{in}|$  & $\frac{I^{+}}{I^{-}}>10$  & $\frac{I^{+}}{I^{-}}>50$  & $\frac{I^{+}}{I^{-}}>10$  & $\frac{I^{+}}{I^{-}}>50$ \tabularnewline
\hline 
100\%  & 343 & 265 (77\%)  & 140 (41\%)  & 193 (56\%)  & 101 (29\%)  & 67 (20\%)  & 38 (11\%) \tabularnewline
90\%  & 354 & 229 (65\%)  & 182 (51\%)  & 120 (34\%)  & 50 (14\%)  & 43 (12\%)  & 18 (5\%) \tabularnewline
70\%  & 361 & 245 (68\%)  & 174 (48\%)  & 98 (27\%)  & 24 (7\%)  & 32 (9\%)  & 10 (3\%) \tabularnewline
50\%  & 379 & 277 (73\%)  & 165 (43\%)  & 49 (13\%)  & 19 (5\%)  & 12 (3\%)  & 5 (1.3\%)\tabularnewline
30\%  & 250 & 203 (81\%)  & 117 (47\%)  & 2 (.6\%)  & -  & - & - \tabularnewline
TOTAL  & 1687 & 1219 (72\%)  & 778 (46\%)  & 462 (26\%)  & 194 (11\%)  & 154 (9\%)  & 71 (4\%) \tabularnewline
\hline 
\end{tabular}

\protect\caption{\label{tab:reduced_contact_fun}\textbf{Functional quality assessment
of models based on randomly reduced contact maps.} The table accounts
only for \emph{conducting} models (1\,pS$<$G$<$1\,nS). The CMAP
column indicates completeness of randomly reduced contact maps.}
\end{table}

\end{doublespace}

\begin{doublespace}

\subsection*{\newpage{}}
\end{doublespace}

\begin{doublespace}
\begin{table}[H]
\begin{tabular}{lccccccc}
\hline 
Condition & RMSD C$_{\alpha}$-C$_{\beta}$ & $Sp\cdot Sn$ & $Sp$  & $Sn$  & $ACC$  & $MCC$  & \emph{$AUROC$}\tabularnewline
\hline 
$\Delta I_{in}<0.80$\,pA  & $<1$\textit{~}Å & 0.37 & 0.52 & 0.71 & 0.60 & 0.23 & 0.65\tabularnewline
$\Delta I_{in}<1.15$\,pA  & $<3$\textit{~}Å & 0.48 & 0.59 & 0.82 & 0.78 & 0.34 & 0.75\tabularnewline
\hline 
$\Delta I_{out}<1.00$\,pA  & $<1$\textit{~}Å & 0.32 & 0.44 & 0.71 & 0.56 & 0.16 & 0.59\tabularnewline
$\Delta I_{out}<1.25$\,pA  & $<3$\textit{~}Å & 0.44 & 0.55 & 0.80 & 0.76 & 0.29 & 0.70\tabularnewline
\hline 
$\Delta\frac{I_{in}^{+}}{I_{in}^{-}}<3.90$ & $<1$\textit{~}Å & 0.48 & 0.69 & 0.69 & 0.69 & 0.38 & 0.72\tabularnewline
$\Delta\frac{I_{in}^{+}}{I_{in}^{-}}<4.40$ & $<3$\textit{~}Å & 0.54 & 0.75 & 0.72 & 0.72 & 0.36 & 0.76\tabularnewline
\hline 
$\Delta\frac{I_{out}^{+}}{I_{out}^{-}}<3.35$ & $<1$\textit{~}Å & 0.48 & 0.67 & 0.73 & 0.69 & 0.39 & 0.74\tabularnewline
$\Delta\frac{I_{out}^{+}}{I_{out}^{-}}<3.60$ & $<3$\textit{~}Å & 0.56 & \multirow{1}{*}{0.85} & \multirow{1}{*}{0.67} & \multirow{1}{*}{0.70} & \multirow{1}{*}{0.38} & 0.78\tabularnewline
\hline 
\end{tabular}

\protect\caption{\label{tab:RMSD_dev} \textbf{Optimal classification based on selectivity
deviation and current deviation related to C$_{\alpha}$-C$_{\beta}$
RMSD as the }\textbf{\emph{ground truth}}\textbf{.}Only the \emph{conducting
}models are considered. The RMSD thresholds corresponded to the following
Positive/Negative ratios: 1\,Å:\,752/935, 3\,Å:\,1412/275. }
\end{table}

\end{doublespace}

\begin{doublespace}

\subsection*{\newpage{}}
\end{doublespace}

\begin{doublespace}
\begin{table}[H]
\protect\caption{\label{tab:Quality-enrichment}\textbf{Quality enrichment of the candidate
subsets using several functional criteria.} Enrichment denotes a fraction
of models fulfilling a functional condition which are within a given
structural or electrostatic threshold. Functional parameters were
calculated only for the \emph{conducting} models. \textbf{(top)} The
functional conditions were defined as follows: \emph{cond}: the \emph{conducting
}models or 1\,pS$<$$G$$<$1~nS, C10: $G$$>$10~pS, S10: $G^{+}/G^{-}>10:1$,
S50: $G^{+}/G^{-}>50:1$, RO: $G_{out}/G_{in}>1$. \textbf{(bottom)}
TOP100 denotes the best 100 models fulfilling all conditions given
as an argument.\emph{ }Here, the enrichment is effectively a fraction
of models which belong to the cross-section of a pair of TOP100 rankings.
Note that models 98th to 107th in the C$_{\alpha}$-C$_{\beta}$ RMSD-based
ranking had exactly the same quality (0.749\,Å). }

\begin{tabular}{lc>{\centering}p{1.2cm}>{\centering}p{1.2cm}>{\centering}p{1.4cm}>{\centering}p{1.4cm}>{\centering}p{1.4cm}>{\centering}p{1.4cm}}
\hline 
\multirow{2}{*}{Condition} & \multirow{2}{*}{\#models} & \multirow{2}{1.2cm}{median RMSD} & \multirow{2}{1.2cm}{median RMSE} & \multicolumn{4}{c}{Enrichment {[}\%{]}}\tabularnewline
\cline{5-8} 
 &  &  &  & RMSD $<1$\,Å & RMSD $<3$\,Å & RMSE $<0.3$\,V & RMSE $<0.5$\,V\tabularnewline
\hline 
none & 2158 & 1.14 & 0.35 & 44 & 82 & 38 & 79\tabularnewline
\emph{cond} & 1674 & 1.12 & 0.35 & 45 & 84 & 37 & 79\tabularnewline
C10 & 1213 & 1.15 & 0.34 & 44 & 82 & 38 & 78\tabularnewline
S10 & 458 & 0.80 & 0.28 & 75 & 99 & 57 & 98\tabularnewline
S50 & 191 & 0.78 & 0.26 & 82 & 100 & 72 & 100\tabularnewline
S10 \& RO & 203 & 0.84 & 0.32 & 75 & 100 & 41 & 98\tabularnewline
S50 \& RO & 71 & 0.78 & 0.29 & 75 & 100 & 61 & 100\tabularnewline
\hline 
 &  &  &  & TOP100 

{[}RMSD{]} & TOP100 {[}RMSE{]} & \multicolumn{2}{>{\centering}p{2.8cm}}{TOP100 {[}RMSD\&RMSE{]}}\tabularnewline
\hline 
none & 2158 & 1.14 & 0.35 & 5.0 & 4.6 & \multicolumn{2}{c}{4.6}\tabularnewline
TOP100{[}$\frac{I_{in}^{+}}{I_{in}^{-}}$\&$\frac{I_{out}^{+}}{I_{out}^{-}}${]}  & 100 & 0.78 & 0.25 & 13 & 16 & \multicolumn{2}{c}{22}\tabularnewline
TOP100{[}$\frac{I_{in}^{+}}{I_{in}^{-}}${]}  & 100 & 0.79 & 0.26 & 9.4 & 16 & \multicolumn{2}{c}{19}\tabularnewline
TOP100{[}$\frac{I_{out}^{+}}{I_{out}^{-}}${]}  & 100 & 0.78 & 0.24 & 18 & 19 & \multicolumn{2}{c}{33}\tabularnewline
\hline 
\end{tabular}
\end{table}

\newpage{}

\end{doublespace}

\section*{Supplemental Data}

\subsection*{Supplemental Table 1}

\begin{table}[H]
\begin{tabular}{llrr}
\hline 
Structural feature  & Functional feature  & $\tau$  & p-value \tabularnewline
\hline 
Pore diameter at THR75  & $|I_{out}/I_{in}|$  & 0.23 - 0.24  & $2.4\times10^{-10}$ -- $5.5\times10^{-11}$ \tabularnewline
RMSD of THR75  & $\Delta(I_{out}/I_{in})$  & 0.14 - 0.21  & $6.5\times10^{-5}$ -- $2.9\times10^{-9}$ \tabularnewline
RMSD of PRO83  & $\Delta I_{in}^{-}$  & 0.20  & $2.3\times10^{-8}$ \tabularnewline
General full atom RMSD  & $\Delta I_{in}^{-}$  & 0.19  & $1.9\times10^{-7}$ \tabularnewline
\hline 
\end{tabular} \label{tab:full_contact_corr} 

\,

Most significant Kendall correlations between structural and functional
features based on the full contact map set. Note that side chains
of reconstructed tetramers of KcsA were not perfectly symmetric which
in case of THR75 resulted in a range of $\tau$.
\end{table}

\newpage{}

\subsection*{Supplemental Table 2}

\begin{table}[H]
\begin{tabular}{crrrr}
\hline 
\multirow{2}{*}{Contact map completeness} & \multirow{2}{*}{Number of models} & \multirow{2}{*}{Full atom RMSD$<$4~Å} & \multicolumn{2}{c}{Median RMSD {[}Å{]}}\tabularnewline
\cline{4-5} 
 &  &  & C$_{\alpha}$-C$_{\beta}$ & Full atom\tabularnewline
\hline 
100\% & 430 & 430 (100\%) & 0.76 & 2.39\tabularnewline
90\% & 460 & 460 (100\%) & 0.85 & 2.42\tabularnewline
70\% & 460 & 460 (100\%) & 1.16 & 2.56\tabularnewline
50\% & 465 & 412 (89\%) & 2.04 & 3.30\tabularnewline
30\% & 361 & 5 (1.4\%) & 5.95 & 6.87\tabularnewline
\hline 
30-100\% & 2176 & 1767 (81\%) & 1.14  & 2.61\tabularnewline
\hline 
\end{tabular}\label{tab:reduced_contact_valid} 

\,

Structural quality assessment of models based on randomly reduced
contact maps.\textbf{ }
\end{table}

\newpage{}

\subsection*{Supplemental Table 3}

\begin{table}[H]
\begin{tabular}{lrrrrr}
\hline 
Structural  & \multicolumn{5}{c}{Functional feature}\tabularnewline
RMSD  & $|\Delta I_{in}|$  & $|\Delta\frac{I_{in}^{+}}{I_{in}^{-}}|$  & $|\Delta I_{out}|$  & $|\Delta\frac{I_{out}^{+}}{I_{out}^{-}}|$  & $|\Delta|\frac{I_{out}^{+}}{I_{in}^{+}}||$\tabularnewline
\hline 
general C$_{\alpha}$-C$_{\beta}$  & 0.23  & 0.32  & 0.15  & 0.34  & 0.01 \tabularnewline
general full atom  & 0.24  & 0.31  & 0.13  & 0.32  & 0.02 \tabularnewline
\hline 
LEU40  & \multirow{6}{*}{0.21-0.23 } & \multirow{6}{*}{0.28-0.30 } & \multirow{6}{*}{0.12-0.15 } & \multirow{6}{*}{0.30-0.31 } & \multirow{6}{*}{-0.01-0.03 }\tabularnewline
PRO63  &  &  &  &  & \tabularnewline
GLY79  &  &  &  &  & \tabularnewline
PRO83  &  &  &  &  & \tabularnewline
GLY88  &  &  &  &  & \tabularnewline
GLY104  &  &  &  &  & \tabularnewline
\hline 
O$_{GLY79}$  & \multirow{2}{*}{0.22-0.23 } & \multirow{2}{*}{0.29 } & \multirow{2}{*}{0.12-0.14 } & \multirow{2}{*}{0.31 } & \multirow{2}{*}{0.01-0.02 }\tabularnewline
O$_{ASP80}$  &  &  &  &  & \tabularnewline
\hline 
\end{tabular}\label{tab:reduced_contact_corr}

\,

Structural features whose Kendall's correlation with at least one
functional feature was above 0.30. 
\end{table}

\newpage{}

\subsection*{Supplemental Table 4}

\begin{table}[H]
\begin{tabular}{cccccccc}
\hline 
Condition & RMSE & $Sp\cdot Sn$ & $Sp$  & $Sn$  & $ACC$  & $MCC$  & \emph{AUROC}\tabularnewline
\hline 
$|\Delta\frac{I_{out}^{+}}{I_{out}^{-}}|<3.90$ & $<0.3$ V & 0.43 & 0.72 & 0.60 & 0.65 & 0.31 & 0.70\tabularnewline
$|\Delta\frac{I_{out}^{+}}{I_{out}^{-}}|<4.15$ & $<0.5$ V & 0.51 & 0.78 & 0.65 & 0.68 & 0.35 & 0.75\tabularnewline
\hline 
$|\Delta\frac{I_{in}^{+}}{I_{in}^{-}}|<2.80$ & $<0.3$ V & 0.44 & 0.83 & 0.53 & 0.65 & 0.36 & 0.71\tabularnewline
$|\Delta\frac{I_{in}^{+}}{I_{in}^{-}}|<3.60$ & $<0.5$ V & 0.51 & 0.76 & 0.67 & 0.69 & 0.36 & 0.76\tabularnewline
\hline 
$|\Delta I_{out}|<0.75$\,pA & $<0.3$ V & 0.38 & 0.59 & 0.65 & 0.63 & 0.23 & 0.67\tabularnewline
$|\Delta I_{out}|<1.00$\,pA & $<0.5$ V & 0.44 & 0.59 & 0.75 & 0.72 & 0.30 & 0.72\tabularnewline
\hline 
$|\Delta I_{in}|<0.90$\,pA & $<0.3$ V & 0.35 & 0.55 & 0.64 & 0.61 & 0.19 & 0.61\tabularnewline
$|\Delta I_{in}|<0.90$\,pA & $<0.5$ V & 0.38 & 0.62 & 0.62 & 0.62 & 0.19 & 0.66\tabularnewline
\hline 
\end{tabular}

\,

Optimal classification parameters based on selectivity deviation and
current deviation related to electrostatic RMSE as the \emph{ground
truth}.
\end{table}

\subsection*{\newpage{}}

\subsection*{Supplemental Figure 1}

\begin{figure}[H]
\centering 

\includegraphics[width=120mm]{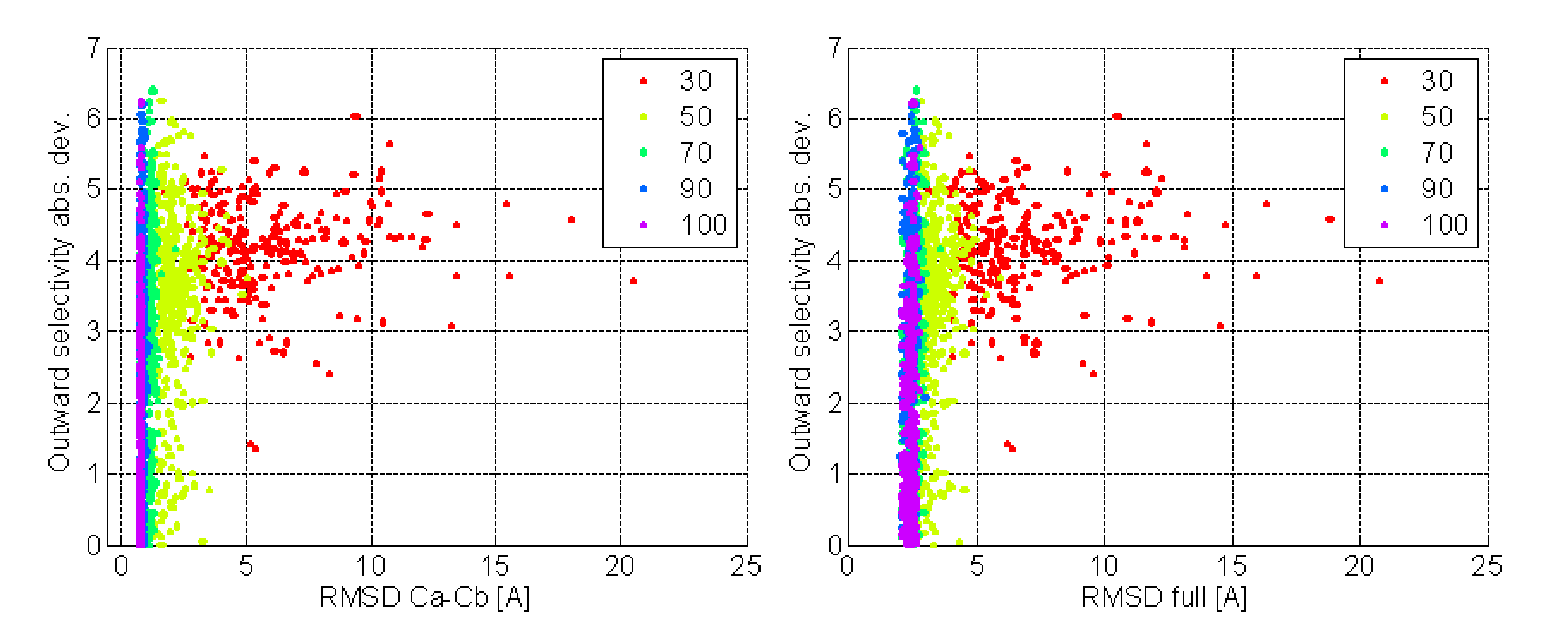}\label{fig:cont30-100_scatter_genrmsd_sel} 

\,

Scatter plot of general RMSD vs. absolute deviation of outward selectivity
\end{figure}

\newpage{}

\subsection*{Supplemental Figure 2}

\begin{figure}[H]
\centering 

\includegraphics[width=120mm]{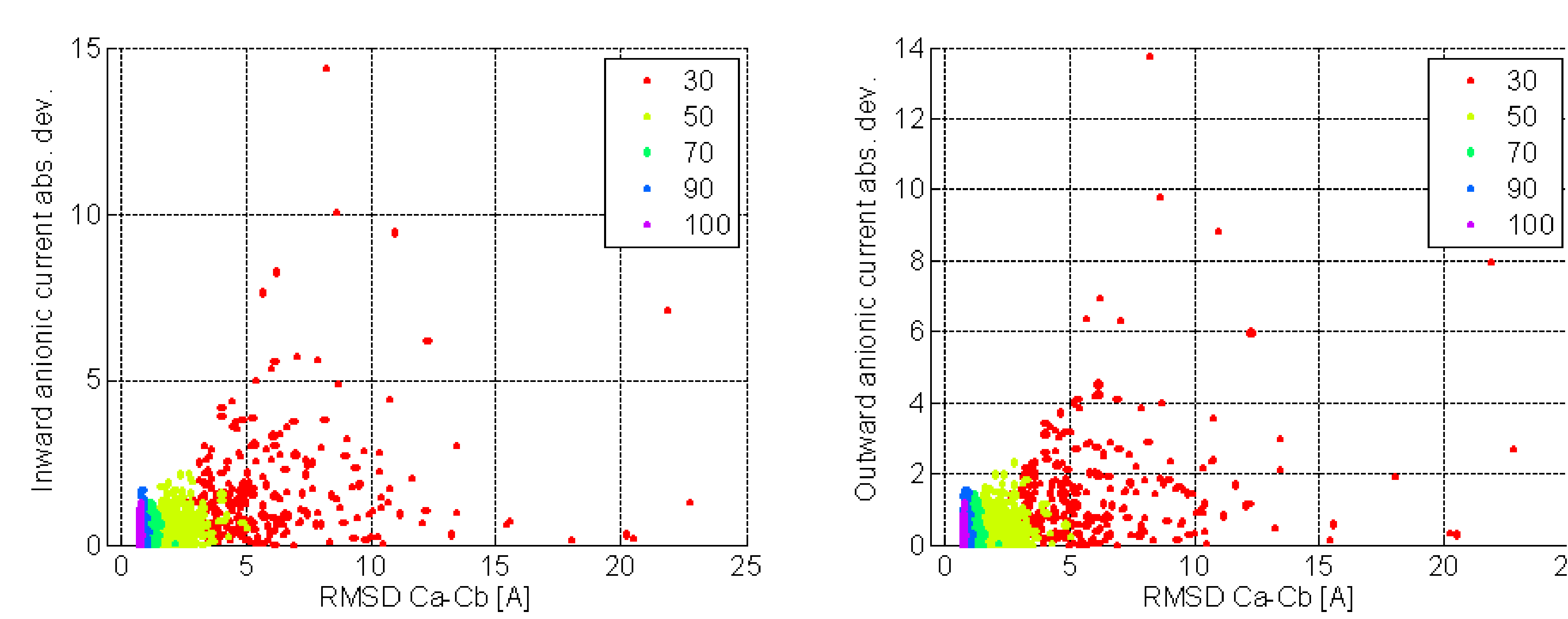}\label{fig:cont30-100_scatter_genrmsd_currani} 

\,

Scatter plot of general C$_{\alpha}$-C$_{\beta}$ RMSD vs. absolute
deviation of anionic current
\end{figure}

\end{document}